\documentclass[12pt,journal,onecolumn]{IEEEtran}
\usepackage{amsmath,amssymb, dsfont,bbm,amsthm}
\usepackage{epsfig,latexsym,amssymb,amsmath,graphics,color}
\usepackage{graphicx,subfigure}
\usepackage{extarrows}
\usepackage{float}
\usepackage[linesnumbered,ruled]{algorithm2e}
\usepackage{epsfig,latexsym,amssymb,amsmath,graphics}
\usepackage{graphicx}
\usepackage[linesnumbered,ruled]{algorithm2e}
\usepackage{cite,color}
\usepackage{url,epstopdf}

\newtheorem{theorem}{Theorem}
\newtheorem{lemma}{Lemma}

\newtheorem{proposition}{Proposition}

\newtheorem{prop}{Proposition}

\newtheorem{cor}{Corollary}

\newtheorem{lm}{Lemma}

\newtheorem{thm}{Theorem}

\newcommand{\be}{\begin{eqnarray}}
\newcommand{\ee}{\end{eqnarray}}
\newcommand{\benn}{\begin{eqnarray*}}
\newcommand{\eenn}{\end{eqnarray*}}
\def\IR{\rm I \kern-0.20em R}

\newcommand{\bthm}{\begin{thm}}
\newcommand{\ethm}{\end{thm}}

\newcommand{\bcor}{\begin{cor}}
\newcommand{\ecor}{\end{cor}}
\newcommand{\bprop}{\begin{prop}}
\newcommand{\eprop}{\end{prop}}
\newcommand{\blm}{\begin{lm}}
\newcommand{\elm}{\end{lm}}
\newcommand{\beq}{\begin{equation}}
\newcommand{\eeq}{\end{equation}}
\newcommand{\ber}{\begin{eqnarray}}
\newcommand{\eer}{\end{eqnarray}}

\newcommand{\bproof}{\begin{proof}}
\newcommand{\eproof}{\end{proof}}

\newcommand{\bit}{\begin{itemize}}
\newcommand{\eit}{\end{itemize}}
\newcommand{\ben}{\begin{enumerate}}
\newcommand{\een}{\end{enumerate}}
\newcommand{\bdesc}{\begin{description}}
\newcommand{\edesc}{\end{description}}
\newcommand{\beqarrn}{\begin{eqnarray*}}
\newcommand{\eeqarrn}{\end{eqnarray*}}
\newcommand{\bproofof}{\begin{proofof}}
\newcommand{\eproofof}{\end{proofof}}
\newenvironment{rem}{\begin{trivlist}\item[]{\bf
Remark:}\hspace{4mm}}{\end{trivlist}}
\newcommand{\brem}{\begin{rem}}
\newcommand{\erem}{\end{rem}}
\newenvironment{rems}{\begin{trivlist}\item[]{\bf
Remarks}\begin{itemize}}{\end{itemize}\end{trivlist}}
\newcommand{\brems}{\begin{rems}}
\newcommand{\erems}{\end{rems}}
\newtheorem{fact}{Fact}
\newcommand{\bfact}{\begin{fact}}
\newcommand{\efact}{\end{fact}}
\newtheorem{examp}{Example}
\newcommand{\bexamp}{\begin{examp}\rm}
\newcommand{\eexamp}{\end{examp}}
\newtheorem{defn}{Definition}
\newcommand{\bdefn}{\begin{defn}\rm}
\newcommand{\edefn}{\end{defn}}

\newtheorem{alg}{Algorithm}
\newcommand{\balg}{\begin{alg}}
\newcommand{\ealg}{\end{alg}}

\newtheorem{prob}{Problem}
\newcommand{\bprob}{\begin{prob}}
\newcommand{\eprob}{\end{prob}}

\newcommand{\bvtm}{\begin{verbatim}}
\newcommand{\bfig}{\begin{figure}}
\newcommand{\efig}{\end{figure}}
\newcommand{\bcen}{\begin{center}}
\newcommand{\ecen}{\end{center}}

\long\def\comment#1{}

\def \n2{{N_0 \over 2}}

\def \h5{\hspace{0.5in}}

\newcommand{\dff}{\stackrel{\triangle}{=}}

\renewcommand{\baselinestretch}{1}

\def\IR{\mathbb R}

\renewcommand{\baselinestretch}{1.6}

\title{On the Sum-Rate Capacity of Poisson Multiple Access Channel with Non-Perfect Photon-Counting Receiver}
\author{Zhimeng Jiang, Chen Gong, Guanchu Wang and Zhengyuan Xu
	\thanks{This work was supported by the National Key Basic Research Program of China (No. 2013CB329201), Key Program of National Natural Science Foundation of China (Grant No. 61631018) and Key Research Program of Frontier Sciences of CAS (Grant No. QYZDY-SSW-JSC003). This paper will be submitted to IEEE Globecom 2019 \cite{jiang2019maccapaconf}.}
	\thanks{The authors are with Key Laboratory of Wireless-Optical Communications, Chinese Academy of Sciences, University of Science and Technology of China, Hefei, Anhui 230027, China. 
		Email: \{zhimengj, hegsns\}@mail.ustc.edu.cn, \{cgong821, xuzy\}@ustc.edu.cn.}}
\date{}

\begin{document}
\maketitle{}

\renewcommand{\baselinestretch}{1.3}
\begin{abstract}
We first investigate two-user nonasymmetric sum-rate Poisson capacity with non-perfect photon-counting receiver under certain condition and demonstrate three possible transmission strategy, including only one active user and both active users, in sharp contrast to Gaussian  multiple access channel (MAC) channel. The two-user capacity reduction due to photon-counting loss is characterized compared with that of continuous Poisson channel. We then study the symmetrical case based on two different methods, demonstrating that the optimal duty cycle for two users must be the same and unique, and the last method maybe can extend to multiple users. Furthermore, we analyze the sum-Rate capacity of Poisson multiple input single output (MISO) MAC. By converting a non-convex optimization problem
with a large number of variables into a non-convex optimization
problem with two variables, we show that the sum-rate capacity
of the Poisson MISO MAC is equivalent to that of SISO under certain condition.
 
\end{abstract}

\renewcommand{\baselinestretch}{1.3}
\section{Introduction}\label{sect.intro}
Due to the potential large bandwidth and no electromagnetic radiation, optical wireless communication shows great promise for the future wireless communications \cite{gagliardi1976optical}. On some specific occasions where the conventional RF is prohibited and direct link transmission cannot be guaranteed, non-line-of-sight (NLOS) optical scattering communication, typically  in the ultra-violet spectrum, can be adopted to guarantee communications requirement \cite{xu2008ultraviolet}. For the NLOS communication, the transmitter and receiver are not required to be perfectly aligned, which
expands the application range beyond the LOS links. Hence, it is difficult to to detect the received signals using a conventional continuous waveform receiver, such as photondiode (PD) and avalanche
photondiode (APD). Instead, a photon-counting receiver is
widely deployed, including photomultiplier tube (PMT) as well as single photon avalanche diode (SPAD).

Poisson channel, whereby the arrival of photons is recorded by photon-sensitive devices incorporated in the receivers, is often used to model free-space optical (FSO) and optical scattering communication. For perfect photon-counting receiver, recent works mainly focus on point-to-point capacity under various scenarios, such as single transmitter \cite{wyner1988capacity,frey1991information}, multiple transmitters \cite{chakraborty2008outage} in continuous-time \cite{shamai1993bounds} and discrete-time
\cite{lapidoth2009capacity,lapidoth2011discrete,wang2014refined}. For multiple users scenario, recent works focus
on the Poisson broadcast channel \cite{lapidoth2003wide}, the Poisson
multiple-access channel (MAC) \cite{lapidoth1998poisson}, and the Poisson interference channel capacity \cite{lai2015capacity}. Moreover, the communication system optimization and corresponding signal processing \cite{el2012binary,jiang2019clipping,gong2015non,ardakani2017performance} have also been extensively studied.

Considering the practical characterization of photon-counting devices, perfect photon-counting receiver assumption is impractical. For example, a typical photon-counting receiver applied in
many optical communication scenarios \cite{chitnis2014spad} includes a photomultiplier tube (PMT) as well as the subsequent sampling and processing blocks \cite{becker2005advanced} or single
photon avalanche diode (SPAD) with quenching circuit.  Specifically, when a photon is captured by receiver, the square pulses subsequently generated by pulse-holding circuits typically have positive width that incurs dead time effect \cite{cherry2012physics}, where a photon arriving during the pulse duration of the previous photon cannot be detected due to the merge of two pulses.The longest time difference of two unrecognized photons is defined as dead time. The photon-counting system with dead time effect for infinite sampling rate and finite sampling rate with shot noise have been investigated in optical wireless communication \cite{sarbazi2018statistical,zou2018characterization}, which shows sub-Poisson distribution of photon counting. In addition, the achievable rate for on-off keying (OOK) modulation and capacity with non-perfect photon-counting receiver are investigated in \cite{jiang2019achievable,jiang2019achievablecapa}. However, The multiuser capacity with non-perfect receiver is still unknown.

In this paper, we investigate the sum-rate capacity of multiple input single output (MISO) multiple access channel (MAC) with non-perfect receiver, assuming negligible electrical thermal noise and shot noise. We first study two-user single-transmitter nonsymmetric sum-rate capacity, which the peak power of two users are not necessary the same, and show that the optimal input signal is two-level piece-wise constant waveform. This scenario naturally arises in multiuser optical communications when the transmitters have different distances to the
receiver or have different transmission powers. Similar to work \cite{lai2017sum}, we resort to Karush-Kuhn-Tucker (KKT) condition to solve the non-convex duty cycle optimization problem and obtain at most four candidate solutions wherein two candidate solutions corresponds
to the cases of only one active user. We further investigate the optimal transmission strategy, corresponding to different possible solutions, for different peak power of each user and give the sufficient condition of each transmission strategy. In particular, we investigate two-user symmetric Poisson channel by two methods based on KKT condition and majorization, both demonstrating that the optimal duty cycle of two users are the same and unique. The last method based on majorization maybe can extend to the case of multiple symmetric users.

We then extend the study to Poisson MAC with multiple
transmitters at each user. Similarly
to the Poisson SISO-MAC, the complex
continuous-input discrete-output Poisson MAC can be converted
to a discrete-time binary-input binary-output Poisson
MAC. However, the joint distribution problem is still challenging since the exponential parameters $2^{J_1}+2^{J_2}$, where $J_1$ and $J_2$ are the total number of transmitters for user $1$ and $2$, respectively. We show that Poisson MISO-MAC capacity equals to Poisson SISO-MAC capacity under certain condition by two steps. The first step is to optimize joint distribution of each user given duty cycle of each transmitter and reduce the dimension from $2^{J_1}+2^{J_2}$ to $J_1+J_2$. The last step is to optimize duty cycle of each transmitter and further reduce the dimension from $J_1+J_2$ to $2$. The key ingredient is to show that,
all antennas at each transmitter being simultaneously on or off achieve the optimality.

The remainder of the paper is organized as follows.
Section~\ref{sect.sysmodel} describes the model under consideration. Section~\ref{sect.sisomac} and Section~\ref{sect.sisomac2}
analyzes the Poisson nonsymmetric and symmetric SISO-MAC capacity, respectively. Section~\ref{sect.misomac} analyzes the Poisson 
Poisson MISO-MAC. Numerical analysis is presented in Section~\ref{sect.numresult}
and concluding remarks are presented in Section~\ref{sect.conclusion}.

\section{System Model}\label{sect.sysmodel}
We introduce the following notations that will be used throughout this paper. Random variables and vectors are denoted by upper-case letters and bold uppercase letters, respectively. We use notation $X_i^j$ to denote a sequence of random variables $\{X_i, X_{i+1},\cdots,X_j\}$; and for $i = 1$, we use $X_{[j]}=\{X_1,\cdots,X_j\}$. A continuous time random process $\{\Lambda(t),a\leq t\leq b\}$ is denoted in short by $\Lambda^b_a$; when $a = 0$, we use $\Lambda^b = \{X(t),0\leq t\leq b\}$. Realizations of random variables and random processes, denoted in lowercase letters, follow the same convention.

Consider multiple users communicating to a single non-perfect receiver. Assume $M$ users, where user $m$, $m=1,\cdots,M$, is equipped with $J_m$ transmitters. Let $\Lambda_{mj}(t)$ denotes the $\mathbb{R}_0^+$-valued photon arrival rate at time $t$ from the $j^{th}$ transmitter of the $m^{th}$ user, and $Y(t)$ denote the Poisson photon arrival process observed at the receiver and
\be 
Y(t)=\mathcal{P}\Big(\sum_{m=1}^{M}\sum_{j=1}^{J_m}\Lambda_{mj}(t)+\Lambda_0\Big),
\ee 
where $\Lambda_0$ is the background radiation, and $\mathcal{P}(\cdot)$ is the Poisson process that records the timing instants and number of photon arrivals. In particular, for any time interval $[t-\tau, t]$, the probability of $k$ photons arriving at the receiver is given by
\be 
\mathbb{P}\{Y(t)-Y(t-\tau)=k\}=\frac{1}{k!}e^{-X_t}(X_t)^k,\quad k=0,1,\cdots,
\ee 
where $X_t=\sum_{m=1}^{M}\sum_{j=1}^{J_m}\int_{t-\tau}^{t}\Lambda_{mj}(t^{'})\mathrm{d}t^{'}$, the arrival rate $\Lambda$ is given by $\Lambda=\frac{P}{h\nu_0}$, where $P$, $h$ and $\nu_0$ denote the transmitted optical power,
the Planck’s constant and the optical spectrum frequency, respectively, such that the energy per photon is given by $h\nu_0$. Thus, the photon arrival rate $\Lambda_{mj}(t)$ must satisfy the following peak power constraint:
\be\label{eq.const1}
0\leq\Lambda_{mj}(t)\leq A_{mj}, 
\ee
where $A_{mj}$ is related to the corresponding maximum power allowed by the $j^{th}$ transmitter of the $m^{th}$ user. In practice,
LEDs or lasers are adopted as the transmitter, where the peak power is limited such that the
peak constraint is more of interest than the average power constraint.

Assuming perfect photon-counting receiver, each photon and the corresponding arrival time can be detected without error. However, perfect photon-counting receiver is difficult to realize and a practical receiver with finite sampling rate consisting of a PMT detector, an one-bit ADC, and a digital signal processor (DSP) unit is considered. When a photon arrives, the PMT detector generates a pulse with certain width, which causes pulses-merge if the interval of two photons arrival time is shorter than the pulse width. The threshold of arrival time interval where the two photons are not differentiable is called dead time, denoted as $\tau$. Denote $T_s$ as the ADC sampling interval and assume low to medium sampling rate such that $T_s\geq\tau$. Considering finite time input $\Lambda_{[MJ]}^T$, the PMT sampling sequence $\mathbf{Z}_{[L]}\dff\{Z_1,\cdots,Z_L\}$, where $L\dff\lfloor\frac{T}{T_s}\rfloor$. Note that for any $\tau>0$, the number of photon arrivals $N_{\tau}$ on
$[0,\tau]$ together with the corresponding (ordered) arrival time instants $\mathbb{T}^{N_{\tau}}=(T_1,\cdots,T_{N_{\tau}})$ are sufficient statistic for $Y^{\tau}$ such that the random vector $(N_T,\mathbb{T}^{N_T})$ is a complete description of random process $Y^T$.  

For the practical photon-counting receiver under consideration, assume zero shot noise, thermal noise and finite dead time. 
For one or multiple photons arriving at the photon-counting receiver at $(iT_s-\tau,iT_s]$, the sampling value $Z_i$ is the same due to the self-sustaining avalanche in SPAD or the shaping circuit that converts bell-shaped response into rectangular response for photon-counting \cite{sarbazi2018statistical,zou2018characterization}. According to above statement, we have 
\be 
Z_i=\left\{\begin{array}{ll}
	0,&T_j\notin(iT_s-\tau,iT_s],\forall j=1,\cdots,N_T; \\
	1,&otherwise;  
\end{array}\right.
\ee 
where $\mathbb{P}(Z_i=1)=1-e^{-X_{iT_s}}$ and $Z_i$ and $Z_j$ are independent identically distributed for $i\neq j$ due to the property of independent increment for Poisson process. In other words, $Z_i$ is an indicator on whether one or more photons arrive within $\tau$ prior to the sampling instant. 

Based on above mentioned system model, the multi-user MISO Poisson channel capacity is defined as
\be
C_{MU-MISO}=\lim\limits_{T\to\infty}\max\limits_{\Lambda_{mj}^T\in[0,A_{mj}]}\frac{1}{T}I(\Lambda_{[MJ]}^T;\mathbf{Z}_{[L]}),
\ee
Since $\Lambda_{[MJ]}^T\rightarrow(N_T,\mathbf{T}^{N_T})\rightarrow\mathbf{Z}_{[n]}$ forms a Markov chain, we have $I(\Lambda_{[MJ]}^T;\mathbf{Z}_{[n]})\leq I(\Lambda_{[MJ]}^T;N_T,\mathbf{T}^{N_T})$, which shows that the multi-user MISO Poisson capacity with non-perfect receiver is not more than that of continuous-time multi-user MISO Poisson channel. 

According to the chain rule for mutual information, we have
\be 
\frac{1}{T}I(\Lambda_{[MJ]}^T;\mathbf{Z}_{[L]})&=&\frac{1}{T}\sum_{l=1}^{L}I(\Lambda_{[MJ],(l-1)T_s}^{lT_s};Z_{l}|\Lambda_{[MJ]}^{(l-1)T_s};\mathbf{Z}_{[l-1]})\nonumber\\
&=&\frac{1}{T}\sum_{l=1}^{L}H(Z_{l}|\Lambda_{[MJ]}^{(l-1)T_s};\mathbf{Z}_{[l-1]})-H(Z_{l}|\Lambda_{[MJ]}^{lT_s};\mathbf{Z}_{[l-1]})\nonumber\\
&\overset{(a)}{=}&\frac{1}{T}\sum_{l=1}^{L}H(Z_{l}|\Lambda_{[MJ]}^{(l-1)T_s};\mathbf{Z}_{[l-1]})-H(Z_{l}|\Lambda_{[MJ],(l-1)T_s}^{lT_s})\nonumber\\
&\leq&\frac{1}{T}\sum_{l=1}^{L}H(Z_{l})-H(Z_{l}|\Lambda_{[MJ],(l-1)T_s}^{lT_s})=\frac{1}{T}\sum_{l=1}^{L}I(\Lambda_{[MJ],(l-1)T_s}^{lT_s};Z_{l}).
\ee 
Where equality (a) holds since $Z_{l}$ is conditional independent of $(\Lambda_{[MJ]}^{(l-1)T_s};\mathbf{Z}_{[l-1]})$ given $\Lambda_{[MJ],(l-1)T_s}^{lT_s}$. Thus, we have 
$C_{MU-MISO}\leq\max\limits_{\Lambda_{mj}^{T_s}\in[0,A_{mj}]}\frac{1}{T_s}I(\Lambda_{[MJ]}^{T_s};\mathbf{Z}_{1}),$
where the equality holds if $\Lambda_{[MJ],(l-1)T_s}^{lT_s}$ is dependent of each other for different $l$. Consequently, the capacity-achieving distribution
requires independent input signals for different sampling intervals, and the simplified capacity
is given by, 
\be\label{eq.capacall}
C_{MU-MISO}=\max\limits_{\Lambda_{mj}^{T_s}\in[0,A_{mj}]}\frac{1}{T_s}I(\Lambda_{[MJ]}^{T_s};\mathbf{Z}_{1}).
\ee 
In the remainder of this paper, we omit subscript $l$ for simplicity since we focus the achievable
rate within a symbol duration to obtain the exact capacity.

\section{SISO Capacity for Two Users}\label{sect.sisomac}
We focus on the case where each user has only one transmitter, i.e., $J_1 = 1$ and $J_2 = 1$. Hence for the sake of convenience, we drop subscript $j$ and use abbreviation $p_1=p(A_1+A_2+\Lambda_0)$, $p_2=p(A_2+\Lambda_0)$, $p_3=p(A_1+\Lambda_0)$, and $p_4=p(\Lambda_0)$. 
\subsection{Optimality Conditions}
The sum-rate capacity is defined as $C_{SISO-MAC}\dff\max\limits_{\Lambda^{\tau}_{m}\in[0,A]}\frac{1}{T_s}I(\Lambda^{T_s}_{[M]};Z)$. The following results show that the optimal
distributions belongs to binary signal levels. 
\begin{theorem}\label{theo.sisomacbi}
	The sum-rate capacity of a Poisson MAC with non-perfect receiver is achieved if the input signal belongs to the set $\{0,A_m\}$ for each user $m$.
	\begin{proof}
		Please refer to Appendix \ref{appen.sisomacbi}.
	\end{proof}
\end{theorem}

Although focusing two-users MAC channel, Theorem \ref{theo.sisomacbi} can be extended to scenario of multiple users. Let $\mu_m$ be the duty cycle of the $m^{th}$ transmitter, $m=1,2$. The sum-rate Poisson MAC capacity is given by
\be\label{eq.sisomaccapa}
C_{SISO-MAC}=\max\limits_{0\leq\mu_1,\mu_2\leq1}\frac{1}{\tau}I_{X_1^2;Z}(\mu_1,\mu_2),
\ee 
where 
\be
I_{X_1^2;Z}(\mu_1,\mu_2)&=&h_b(\hat{p}(\mu_1,\mu_2))-\mu_1\mu_2h_b(p_1)-(1-\mu_1)\mu_2h_b(p_2)\nonumber\\&&-\mu_1(1-\mu_2)h_b(p_3)-(1-\mu_1)(1-\mu_2)h_b(p_4),\\
\hat{p}(\mu_1,\mu_2)&=&\mu_1\mu_2p_1+(1-\mu_1)\mu_2p_2+\mu_1(1-\mu_2)p_3+(1-\mu_1)(1-\mu_2)p_4,
\ee
For the problem (\ref{eq.sisomaccapa}), we have the following property.
\begin{lemma}\label{lemmal.sisomacnonconacve}
	Assume that $\tau\leq\frac{\ln2}{A_1+A_2+\Lambda_0}$. For general values of $A_1,A_2$ and $\Lambda_0$, $I_{X_1^2;Z}(\mu_1,\mu_2)$ is not necessarily a concave function of $(\mu_1,\mu_2)$. In addition, the optimized joint distribution set is not convex.
	\begin{proof}
		Please refer to Appendix \ref{appen.sisomacconcave}.
	\end{proof}
\end{lemma}
According to Lemma \ref{lemmal.sisomacnonconacve}, Problem (\ref{eq.sisomaccapa}) is a non-convex optimization problem in general.

We focus on solving such non-convex optimization problem. We start with the necessary KKT conditions (since the problem is not convex, these conditions are not sufficient
for optimality). For convenience, we write $I_{X_1^2;Z}=I$, and thus the corresponding Lagrangian equation is given by,
\be 
\mathcal{L}=-I+\eta_1(\mu_1-1)-\eta_2\mu_1+\eta_3(\mu_2-1)-\eta_4\mu_2.
\ee 
The optimal solution $(\hat{\mu}_1,\hat{\mu}_2)$ must satisfy the following KKT
constraints:
\be
\frac{\partial I}{\partial \mu_1}|_{(\hat{\mu}_1,\hat{\mu}_2)}-\eta_1+\eta_2=0,\label{eq.sisomackkt1}\\
\frac{\partial I}{\partial \mu_2}|_{(\hat{\mu}_1,\hat{\mu}_2)}-\eta_3+\eta_4=0,\label{eq.sisomackkt2}\\
\eta_1(\hat{\mu}_1-1)=0,
\eta_2\hat{\mu}_1=0,
\eta_3(\hat{\mu}_2-1)=0,
\eta_4\hat{\mu}_2=0,\nonumber
\ee
where
\be 
\frac{\partial I}{\partial \mu_1}&=&\big[\mu_2(p_1-p_2)+(1-\mu_2)(p_3-p_4)\big]\ln\frac{1-\hat{p}}{\hat{p}}\nonumber\\&&-\big[\mu_2\big(h_b(p_1)-h_b(p_2)\big)+(1-\mu_2)\big(h_b(p_3)-h_b(p_4)\big)\big],\label{eq.firstdI1}\\
\frac{\partial I}{\partial \mu_2}&=&\big[\mu_1(p_1-p_3)+(1-\mu_1)(p_2-p_4)\big]\ln\frac{1-\hat{p}}{\hat{p}}\label{eq.firstdI2}\nonumber\\&&-\big[\mu_1\big(h_b(p_1)-h_b(p_3)\big)+(1-\mu_1)\big(h_b(p_2)-h_b(p_4)\big)\big],
\ee

Note that $\eta_1\eta_2=0$ and $\eta_3\eta_4=0$, in order to further analyze the above KKT conditions, we need to consider $5$ cases corresponding to different combinations of active constraints. Similar to work \cite{lai2017sum}, we can show that two cases are non-optimal solutions. For example, if $\eta_1=0$, $\eta_2=0$, $\eta_3\neq0$, $\eta_4=0$, we have $I(\bar{\mu}_1,1)<I(\bar{\mu}_1,0)$ where $I(\bar{\mu}_1,0)$ corresponds to \textit{Scenario 2}. Therefore, the rest three possible scenarios need further investigation.

\textit{Scenario 1:} $\eta_1=0$, $\eta_2=0$, $\eta_3=0$, and $\eta_4=0$.

The KKT conditions can be simplified to $\frac{\partial I}{\partial \mu_1}|_{(\hat{\mu}_1,\hat{\mu}_2)}=0$ and $\frac{\partial I}{\partial \mu_2}|_{(\hat{\mu}_1,\hat{\mu}_2)}=0$. This scenario corresponds to the case where both users are active. As both $\frac{\partial I}{\partial \mu_1}|_{(\hat{\mu}_1,\hat{\mu}_2)}$ and $\frac{\partial I}{\partial \mu_2}|_{(\hat{\mu}_1,\hat{\mu}_2)}$ are nonlinear,
there can be multiple possible $(\mu_1,\mu_2)$ pairs solution. However, we now show that there are at most $2$ possible $(\mu_1,\mu_2)$ pairs solution. By removing the common item $\ln\frac{1-\hat{p}}{\hat{p}}$ in equations (\ref{eq.sisomackkt1}) and (\ref{eq.sisomackkt2}), we have
\be\label{eq.sisomacline} 
&&\mu_1U-\mu_2V+W=0,
\ee
where 
\be 
U&=&-h_{13}+h_{14}+h_{23}-h_{24}+h_{31}-h_{32}-h_{41}+h_{42},\\ V&=&-h_{12}+h_{14}+h_{21}-h_{23}+h_{32}-h_{34}-h_{41}+h_{43},\\ W&=&-h_{23}+h_{24}+h_{32}-h_{34}-h_{42}+h_{43},
\ee
and $h_{ij}\dff p_ih_b\big(p_j\big)$ for $i,j=1,2,3,4$. Regarding equation (\ref{eq.sisomacline}), we have the following property on $U,V,W$.
\begin{lemma}\label{lemma.pnuvw}
	For any $A_1,A_2,\Lambda_0$, we have $U>0$ and $V>0$; and we have $W\lesseqqgtr0$ if and only if $A_1\gtreqqless A_2$.
	\begin{proof}
		Please refer to Appendix \ref{appen.pnuvw}.
	\end{proof}
\end{lemma}

Based on equation (\ref{eq.sisomacline}) and Lemma \ref{lemma.pnuvw}, we have
\be\label{eq.sisomacf}
\mu_2=\frac{U}{V}\mu_1+\frac{W}{V}\dff f_{MAC}(\mu_1).
\ee 
As $\frac{\partial I}{\partial \mu_2}=0$, we have
\be \label{eq.sisomacg}
\mu_2=\frac{(a_M+1)^{-1}-[\mu_1p_3+(1-\mu_1)p_4]}{\mu_1(p_1-p_3)+(1-\mu_1)(p_2-p_4)}\dff g_{MAC}(\mu_1),
\ee 
where $a_M=\exp\Big(\frac{\mu_1\big(h_b(p_1)-h_b(p_3)\big)+(1-\mu_1)\big(h_b(p_2)-h_b(p_4)\big)}{\mu_1(p_1-p_3)+(1-\mu_1)(p_2-p_4)}\Big)$. Hence, the $(\mu_1,\mu_2)$ pairs where $f_{MAC}(\mu_1)$ and $g_{MAC}(\mu_1)$ intersect with each other satisfy equations (\ref{eq.sisomacf}) and (\ref{eq.sisomacg}) simultaneously.

For function $g_{MAC}(\mu_1)$, we have the following lemma \ref{lemma.sisomacconvexg}.
\begin{lemma}\label{lemma.sisomacconvexg}
	Assume $\tau\leq\frac{\ln2}{A_1+A_2+\Lambda_0}$. Then, $g_{MAC}(\mu_1)$ is a strictly convex function with respect to $\mu_1$.
	\begin{proof}
		Please refer to Appendix \ref{appen.convexgmac}.
	\end{proof}
\end{lemma}

As $f_{MAC}(\mu_1)$ is a linear with respect to $\mu_1$, and $g_{MAC}(\mu_1)$ is a strictly convex function with respect to $\mu_1$ , there will be at most two intersection points, denoted as $(\hat{\mu}_1,\hat{\mu}_2)$ and $(\check{\mu}_1,\check{\mu}_2)$. We then need to check whether
$(\hat{\mu_1},\hat{\mu_2})$ is in $[0,1]^2$ or not. If yes, we keep it. If not,
then for the presentation convenience, we replace it with $(0,0)$.

\textit{Scenario 2:} $\eta_1=0$,$\eta_2=0$,$\eta_3=0$, and $\eta_4\neq0$.

Solving the corresponding KKT conditions, we obtain $\tilde{\mu}_1=\alpha_{\tau}(A_1,\Lambda_0)$ and $\tilde{\mu}_2=0$, where 
\be 
\alpha_{\tau}(A_1,\Lambda_0)=\frac{[1+\exp(\frac{h_b(p_3)-h_b(p_4)}{p_3-p_4})]^{-1}-p_4}{p_3-p_4},
\ee 
It is seen that $0 \leq \alpha_{\tau}(A_1, \Lambda_0) \leq 1$, since
\be p_4=[1+\exp(h_b^{'}(p_4))]^{-1}\leq[1+\exp(\frac{h_b(p_3)-h_b(p_4)}{p_3-p_4})]^{-1}\leq[1+\exp(h_b^{'}(p_3))]^{-1}=p_3.\nonumber
\ee
This scenario corresponds to the case where only user $1$ is active.

\textit{Scenario 3:} $\eta_1=0$, $\eta_2\neq0$, $\eta_3=0$, and $\eta_4=0$.

Solving the corresponding KKT conditions, we obtain $\bar{\mu}_2=\alpha_{\tau}(A_2,\Lambda_0)$ and $\bar{\mu}_1=0$, where 
\be 
\alpha_{\tau}(A_2,\Lambda_0)=\frac{[1+\exp(\frac{h_b(p_2)-h_b(p_4)}{p_2-p_4})]^{-1}-p_4}{p_2-p_4}.
\ee 
It is seen that $0 \leq \alpha_{\tau}(A_2, \Lambda_0) \leq 1$, since
\be p_4=[1+\exp(h_b^{'}(p_4))]^{-1}\leq[1+\exp(\frac{h_b(p_2)-h_b(p_4)}{p_2-p_4})]^{-1}\leq[1+\exp(h_b^{'}(p_2))]^{-1}=p_2.\nonumber
\ee
This scenario corresponds to the case where only user $2$ is active.

In summary, we have the following theorem.
\begin{theorem}\label{theo.maccapa}
	Assume that $\tau\leq\frac{\ln2}{A_1+A_2+\Lambda_0}$. The sum-rate capacity of the Poisson MAC is given by
	\be C_{SISO-MAC}=\frac{1}{\tau}\max\{I_{X_1^2;Z}(\hat{\mu}_1,\hat{\mu}_2),I_{X_1^2;Z}(\check{\mu}_1,\check{\mu}_2),I_{X_1^2;Z}(\tilde{\mu}_1,\tilde{\mu}_2),I_{X_1^2;Z}(\bar{\mu}_1,\bar{\mu}_2)\}.
	\ee
\end{theorem}

Unlike the Gaussian MAC with an average power constraint, it can be optimal to allow only one user to transmit in order to achieve the sum-rate capacity
for the Poisson MAC with a peak power constraint. More detailed discussions are presented in the following subsection.

\subsection{Single-User or Two-User Transmission?}\label{subsect.stragety}
We present sufficient conditions on the optimality of a single-user transmission and two-user transmission.

Similar to work \cite{lai2017sum}, the following simple proposition characterize the sufficient conditions where there is no intersection between equations (\ref{eq.sisomacf}) and (\ref{eq.sisomacf}) in duty cycle feasible region $[0,1]^2$ and hence two-user transmission is not optimal.
\begin{proposition}\label{prop.sisomacsingt}
	If $g_{MAC}(0)<f_{MAC}(0)$ and $g_{MAC}(1)<f_{MAC}(1)$, then
	single-user transmission is optimal to achieve the sum-rate
	capacity.
\end{proposition}

Even if the sufficient conditions in Proposition \ref{prop.sisomacsingt} are not satisfied, it is still possible for single-user transmission to be optimal if the corresponding rate
is larger than that of the two-user transmission. We conclude that if certain $A_m$ is sufficiently large, single-user transmission is optimal.
\begin{lemma}\label{lemma.sisomacstra1}
	Functions $f_{MAC}(\mu_1)$ and $g_{MAC}(\mu_1)$ have the
	following properties:
	\be 
	\lim\limits_{A_2\to\infty}f_{MAC}(\mu_1)=\lim\limits_{A_2\to\infty}f_{MAC}(0)=\lim\limits_{A_2\to\infty}f_{MAC}(1)=1,\\
	\lim\limits_{A_2\to\infty}g_{MAC}(\mu_1)=\frac{(a_{MI}+1)^{-1}-[\mu_1p_3+(1-\mu_1)p_4]}{\mu_1(p_1-p_3)+(1-\mu_1)(p_2-p_4)}<1,\text{ for any }\mu_1\in[0,1],
	\ee 
	where $a_{MI}=\exp(-\frac{\mu_1h_b(p_3)+(1-\mu_1)h_b(p_4)}{\mu_1p_3+(1-\mu_1)p_4})$.
	\begin{proof}
		Please refer to Appendix \ref{appen.sisomacstra1}.
	\end{proof}
\end{lemma}

Lemma \ref{lemma.sisomacstra1} and Proposition \ref{prop.sisomacsingt} imply that a single active user is optimal for sufficient high peak power constraint of other user given peak power constraint of certain user. Furthermore, it is seen that the sum-rate capacity is achieved when only user $2$ is transmitting.

We further discuss the conditions on the optimality of two-user transmission. The following proposition characterizes sufficient conditions where single-user transmission is not optimal.
\begin{proposition}\label{prop.sisomacnec}
	Single user $1$ transmission is not optimal if $\frac{\partial I}{\partial \mu_2}|_{(\tilde{\mu}_1,0)}>0$. Similarly, single user $2$ transmission
	alone is not optimal if $\frac{\partial I}{\partial \mu_1}|_{(0,\bar{\mu}_2)}>0$.
	\begin{proof}
		Please refer to Appendix \ref{appen.sisomacnec}.
	\end{proof}
\end{proposition}

\subsection{Asymptotic Capacity Property for $\tau\to0$}
We further investigate the asymptotic properties of the non-perfect receiver compared with the continuous Poisson channel, summarized in Theorem~\ref{theo.macasymp}. The main clue is to show the asymptotic properties of optimized objective function and optimal duty cycle.

\begin{theorem}\label{theo.macasymp}
	The optimal duty cycle and MAC capacity of the non-perfect receiver approach those of continuous Poisson channel for any bounded $A_1$, $A_2$ and $\Lambda_0$, respectively, as $\tau\to0$.
	\begin{proof}
		Please refer to Appendix \ref{appen.macasymp}.
	\end{proof}
\end{theorem}

Theorem~\ref{theo.macasymp} studies the asymptotic property of the non-perfect receiver for $T_s=\tau\to0$. It
shows that Theorem~\ref{theo.maccapa} extends the result of continuous MAC Poisson capacity \cite{lai2017sum}, and provides a
more general and practical results.

\section{SISO Capacity for Symmetric Two Users}\label{sect.sisomac2}
Section~\ref{sect.sisomac} demonstrates SISO capacity for general two users based on KKT conditions. However, this method is hard to extend for multiple users since exponential number of Lagrangian multipliers. In this Section, we reduce the number of candidate optimal solutions from $4$ to $1$ for symmetric channel based on Section~\ref{sect.sisomac}, and provide another method to find optimal solution based on majorization. The notation in this section is similar to Section~\ref{sect.sisomac} and $p_2=p_3$ for symmetric channel.
\subsection{KKT Conditions Perspective}\label{subsect.KKT}
For symmetric channel, we prove that the optimal transmission strategy is two-user transmission with the same and unique duty cycle. The proof is given by the following three
steps.

\textit{Step 1:} We prove that two-user transmission is the optimal transmission strategy for $A_1=A_2$. When $(\mu_1,\mu_2)=(\tilde{\mu}_1,0)$, we have 
\be
\hat{p}&=&\tilde{\mu}_1p_3+(1-\tilde{\mu}_1)p_4=[1+\exp(\frac{h_b(p_3)-h_b(p_4)}{p_3-p_4})]^{-1},
\ee 
Note that $\frac{h_b(p_1)-h_b(p_3)}{p_1-p_3}<\frac{h_b(p_2)-h_b(p_4)}{p_2-p_4}$, according to lemma \ref{appenAu.sisomacaux}, we have
\be 
\frac{\mu_1\big(h_b(p_1)-h_b(p_3)\big)+(1-\mu_1)\big(h_b(p_2)-h_b(p_4)\big)}{\mu_1(p_1-p_3)+(1-\mu_1)(p_2-p_4)}\leq\frac{h_b(p_2)-h_b(p_4)}{p_2-p_4}=\ln\frac{1-\hat{p}}{\hat{p}},
\ee 
which implies $\frac{\partial I}{\partial \mu_2}|_{(0,\bar{\mu}_2)}>0$. Thus, single active user $1$ is not optimal. Similarly, single active user $2$ is not optimal.

\textit{Step 2:} We prove that $\mu_1=\mu_2$ is optimal for both active users. Note that in such a scenario, $p_2=p_3$, $h_{2\cdot}=h_{3\cdot}$ and $h_{\cdot2}=h_{\cdot3}$. Thus, we have $W=0$ and $U=-V$, i.e., for the optimal $(\mu_1,\mu_2)$ we have $\mu_1=\mu_2$.

\textit{Step 3:} We finally prove that there exists unique pair $(\mu_1,\mu_2)$ that satisfies equation (\ref{eq.sisomacf}) and (\ref{eq.sisomacg}). It is easy to check that $g_{MAC}(0)>0=f_{MAC}(0)$ and $g_{MAC}(1)<1=f_{MAC}(1)$. Thus, there exists a single intersection between $f_{MAC}(\mu)$ and $g_{MAC}(\mu)$ for $0\leq\mu_1\leq1$.

\subsection{Majorization Perspective}\label{subsect.majorization}
KKT-conditions-based method provides the necessary condition for the optimal solution, but it is hard to capture the specific property for the objective function and extend to multiple users. We investigate problem (\ref{eq.sisomaccapa}) based on majorization and obtain the same result as Section~\ref{subsect.KKT}. In addition, majorization-based method reveals more information about the problem (\ref{eq.sisomaccapa}) and maybe can be extended to the scenario of multiple users.

Recall the sum-rate Poisson MAC capacity $C_{SISO-MAC}=\max\limits_{0\leq\mu_1,\mu_2\leq1}\frac{1}{\tau}I_{X_1^2;Z}(\mu_1,\mu_2)$,
where $I_{X_1^2;Z}(\mu_1,\mu_2)=h_b(\hat{p}(\mu_1,\mu_2))-\mu_1\mu_2h_b(p_1)-(1-\mu_1)\mu_2h_b(p_2)-\mu_1(1-\mu_2)h_b(p_3)-(1-\mu_1)(1-\mu_2)h_b(p_4)$, $\hat{p}(\mu_1,\mu_2)=\mu_1\mu_2p_1+(1-\mu_1)\mu_2p_2+\mu_1(1-\mu_2)p_3+(1-\mu_1)(1-\mu_2)p_4$ and $p_2=p_3$. The solution based on majorization consists the following two steps corresponding to the inner and outer optimization as $C_{SISO-MAC}=\max\limits_{0\leq\mu_s\leq1}\frac{1}{\tau}I_2(\mu_s)$, where $I_2(\mu_s)=\max\limits_{{\begin{subarray}{c}(\mu_1,\mu_2):\\ \mu_1+\mu_2=2\mu_s\end{subarray}}}I_{X_1^2;Z}(\mu_1,\mu_2)$.

\textit{Step 1:} Assume that $\tau\leq\frac{\ln2}{2A+\Lambda_0}$. We optimize $\mu_1$ and $\mu_2$ with the constraint $\mu_1+\mu_2=2\mu_{s}$ for any given $0\leq\mu_{s}\leq 1$.

Firstly, we provide two critical Lemmas as follows,
\begin{lemma}\label{lemma.sisomacsymLem}
	Assume that $\tau\leq\frac{\ln2}{2A+\Lambda_0}$. Define $G(A)=\frac{2h_b(p_2)-h_b(p_1)-h_b(p_4)}{2p_2-p_1-p_4}$, then we have $G(A)$ decreases with peak power $A$ and $G(A)\in(\ln(1-p_4)+\frac{p_4}{1-p_4}\ln p_4,\frac{1}{p_4}+\ln\frac{1-p_4}{p_4})$.
	\begin{proof}
		Please refer to Appendix~\ref{appen.sisomacsymLem}.
	\end{proof}
\end{lemma}
\begin{lemma}\label{lemma.sisomacsymLem2}
	The solution $\ln\frac{1-\hat{p}}{\hat{p}}=G(A)$ for $(\mu_1,\mu_2)\in[0,1]^2$ iff $A\geq A_{th}$, where $A_{th}$ is the unique solution on $\ln\frac{1-p_4}{p_4}=G(A)$.
	\begin{proof}
		Please refer to Appendix~\ref{appen.sisomacsymLem2}.
	\end{proof}
\end{lemma} 

We focus on the region $\mu_1\geq\mu_2$ since the objective function in Equation (\ref{eq.sisomaccapa}) and the feasible region are symmetric for $\mu_1$ and $\mu_2$. Based on Equations (\ref{eq.firstdI1}) and (\ref{eq.firstdI2}), we have
\be \label{eq.majori1}
\frac{\partial I}{\partial \mu_1}-\frac{\partial I}{\partial \mu_2}&=&\Big\{[\mu_2(p_1-p_2)+(1-\mu_2)(p_3-p_4)]\ln\frac{1-\hat{p}}{\hat{p}}-[\mu_2\big(h_b(p_1)-h_b(p_2)\big)\nonumber\\&&+(1-\mu_2)\big(h_b(p_3)-h_b(p_4)\big)]\Big\}-\Big\{[\mu_1(p_1-p_3)+(1-\mu_1)(p_2-p_4)]\ln\frac{1-\hat{p}}{\hat{p}}\nonumber\\
&&-[\mu_1\big(h_b(p_1)-h_b(p_3)\big)+(1-\mu_1)\big(h_b(p_2)-h_b(p_4)\big)]\Big\}\nonumber\\
&\xlongequal{p_2=p_3}&(\mu_1-\mu_2)\Big\{\ln\frac{1-\hat{p}}{\hat{p}}(2p_2-p_1-p_4)-\big[2h_b(p_2)-h_b(p_1)-h_b(p_4)\big]\Big\}\nonumber\\
&=&(\mu_1-\mu_2)(2p_2-p_1-p_4)\big\{\ln\frac{1-\hat{p}}{\hat{p}}-G(A)\big\}
\ee 

According to Lemma~\ref{lemma.sisomacsymLem2}, we can analysis Equation (\ref{eq.majori1}) by two cases.

\textbf{Case 1:} $A<A_{th}$. According to Lemma~\ref{lemma.sisomacsymLem2}, we have $\ln\frac{1-\hat{p}}{\hat{p}}-G(A)<0$ and $(\mu_1-\mu_2)(\frac{\partial I}{\partial \mu_1}-\frac{\partial I}{\partial \mu_2})<0$ for $(\mu_1,\mu_2)\in[0,1]^2$. According to \cite[A.4. Theorem, p.84]{marshall1979inequalities}, we have that mapping $(\mu_1,\mu_2)\longmapsto I_{X_1^2;Z}(\mu_1,\mu_2)$ is Schur-concave for $(\mu_1,\mu_2)\in[0,1]^2$ and the optimal $(\mu_1,\mu_2)$ with the constraint $\mu_1+\mu_2=2\mu_{s}$ is $(\mu_s,\mu_s)$.

\textbf{Case 2:} $A\geq A_{th}$. Define $\mathcal{C}\dff\{(\mu_1,\mu_2):\ln\frac{1-\hat{p}}{\hat{p}}=G(A),\mu_1\geq\mu_2\}$ and $\mathcal{L}_{\mu_s}\dff\{(\mu_1,\mu_2):\mu_1+\mu_2=2\mu_s,\mu_1\geq\mu_2\}$. According to Lemma~\ref{lemma.sisomacsymLem2}, we have $\mathcal{C}\neq\varnothing$. We further investigate the property of $\mathcal{C}$ as shown in Theorem~\ref{theo.sisomacsymBound}.

\begin{theorem}\label{theo.sisomacsymBound}
	Assume that $\tau\leq\frac{\ln2}{2A+\Lambda_0}$. There exists differentiable function $f_B(\cdot)$ such that $\mathcal{C}=\{(\mu_1,\mu_2):\mu_1=f_B(\mu_2)\}$, where $0\geq f_B(0)<1$ and $f_B(\mu_2)<-1$ for $\mu_2\in[0,\frac{1}{2}]$. In addition, $|\mathcal{C}\cap\mathcal{L}_{\mu_s}|=1$ for $\mu_s^{'}\leq\mu_s\leq\mu_s^{*}$, and $|\mathcal{C}\cap\mathcal{L}_{\mu_s}|=0$ for $\mu_s\geq\mu_s^{*}$ and $\mu_s\leq\mu_s^{'}$, where $\mu_s^{'}=\frac{(p_2-p_4)-\sqrt{(p_2-p_4)^2-(2p_2-p_1-p_4)[\frac{1}{1+\exp(G(A))}-p_4]}}{2p_2-p_1-p_4}$ and $\mu_s^{*}=\frac{\frac{1}{1+\exp(G(A))}-p_4}{2(p_2-p_4)}<\frac{\tilde{\mu}_1}{2}$.
	\begin{proof}
		Please refer to Appendix~\ref{appen.sisomacsymBound}.
	\end{proof}
\end{theorem} 

According to Theorem~\ref{theo.sisomacsymBound} and $\hat{p}$ increases with $\mu_1$ and $\mu_2$, define $\mathcal{C}^{+}=\{(\mu_1,\mu_2):\mu_1\geq f_B(\mu_2)\}$ and $\mathcal{C}^{-}=\{(\mu_1,\mu_2):\mu_1< f_B(\mu_2)\}$, we have that that mapping $(\mu_1,\mu_2)\longmapsto I_{X_1^2;Z}(\mu_1,\mu_2)$ is Schur-concave and Schur-convex for region $\mathcal{C}^{+}$ and $\mathcal{C}^{-}$, respectively. Thus, $I_2(\mu_s)$ is given by
\be \label{eq.candidateI2}
I_2(\mu_s)=\left\{\begin{array}{ll}
	I_{X_1^2;Z}(2\mu_s,0),&\mu_s\leq\mu_s^{'}, \\
	\max\{I_{X_1^2;Z}(2\mu_s,0),I_{X_1^2;Z}(\mu_s,\mu_s)\},&\mu_s^{'}<\mu_s<\mu_s^{*}, \\
	I_{X_1^2;Z}(\mu_s,\mu_s),&\mu_s\geq\mu_s^{*}.  
\end{array}\right.
\ee 

\textit{Step 2:} We optimize $\mu_s$ to maximize $I_2(\mu_s)$ over $\mu_s\in[0,1]$.

According to Equation (\ref{eq.candidateI2}), we have the candidate solution to maximize $I_{X_1^2;Z}(\mu_1,\mu_2)$ over $\mu_1\geq\mu_2$ are $(2\mu_s,0)$ for $0\leq\mu_s\leq\mu_s^{*}$, and $(\mu_s,\mu_s)$ for $1\geq\mu_s\geq\mu_s^{'}$. According to $\mu_s^{*}<\frac{\tilde{\mu}_1}{2}$ and \textit{Scenario 2} in Section~\ref{sect.sisomac}, we have $I_{X_1^2;Z}(2\mu_s,0)$ increases with $\mu_s$ over $\mu_s\leq\mu_s^{*}$.
Note that
\be 
\ln\frac{1-\hat{p}(2\mu_s^{*},0)}{\hat{p}(2\mu_s^{*},0)}&=&h_b^{'}(2\mu_s^{*}p_3+(1-2\mu_s^{*})p_4)=G(A)\nonumber\\&\overset{(a)}{>}&\frac{h_b(p_2)-h_b(p_4)}{p_2-p_4}\overset{(b)}{>}\frac{\mu\big(h_b(p_1)-h_b(p_3)\big)+(1-\mu)\big(h_b(p_2)-h_b(p_4)\big)}{\mu(p_1-p_3)+(1-\mu)(p_2-p_4)},
\ee 
where $(a)$ and $(b)$ hold according to Lemma~\ref{appenAu.last} and Lemma~\ref{appenAu.sisomacaux}, respectively. Thus,
we have 
\be 
\frac{\partial I}{\partial \mu_2}\Big|_{(\mu_s^{*},0)}&=&[\mu_s^{*}(p_1-p_3)+(1-\mu_s^{*})(p_2-p_4)]\ln\frac{1-\hat{p}}{\hat{p}}\nonumber\\&&-[\mu_s^{*}\big(h_b(p_1)-h_b(p_3)\big)+(1-\mu_s^{*})\big(h_b(p_2)-h_b(p_4)\big)]>0.
\ee 
and the optimal solution to maximize $I_{X_1^2;Z}(\mu_1,\mu_2)$ is not in $(2\mu_s,0)$ for $0\leq\mu_s\leq\mu_s^{*}$. 

For the rest candidate region $(\mu_s,\mu_s)$ for $1\geq\mu_s\geq\mu_s^{'}$, it is easy to check that $I_{X_1^2;Z}(\mu_s^{'},\mu_s^{'})=I_{X_1^2;Z}(2\mu_s^{*},0)$ and $I_{X_1^2;Z}(1,1)=0$. For the continuous objection function $I_{X_1^2;Z}(\mu_s,\mu_s)$ over $1\geq\mu_s\geq\mu_s^{'}$, the optimal solution must be a extreme point satisfying the following equation,
\be\label{eq.firstdneces} 
0&=&\frac{\partial I_{X_1^2;Z}(\mu_s,\mu_s)}{\partial \mu_s}=(\frac{\partial I_{X_1^2;Z}}{\partial \mu_1}-\frac{\partial I_{X_1^2;Z}}{\partial \mu_2})\Big|_{(\mu_s,\mu_s)}\nonumber\\&=&2\Big\{[\mu_s(p_1-p_2)+(1-\mu_s)(p_2-p_4)]\ln\frac{1-\hat{p}}{\hat{p}}\nonumber\\&&-[\mu_s\big(h_b(p_1)-h_b(p_2)\big)+(1-\mu_s)\big(h_b(p_2)-h_b(p_4)\big)]\big\},
\ee 
where $\hat{p}=\mu_s^2p_1+2\mu_s(1-\mu_s)p_2+(1-\mu_s)^2p_4$. It is easy to check that Equation (\ref{eq.firstdneces}) equals to $\mu_s=g_{MAC}(\mu_s)$ in Equation (\ref{eq.sisomacg}). According to Section~\ref{subsect.KKT}, we have that there exists unique solution on Equation (\ref{eq.firstdneces}). 

Section~\ref{subsect.majorization} shows that the optimal solution to maximize $I_{X_1^2;Z}(\mu_1,\mu_2)$ satisfies $\mu_1=\mu_2=\mu_s$ and $\mu_s$ is the unique solution on Equation (\ref{eq.firstdneces}), the same as the result in Section~\ref{subsect.KKT}. In addition, Work \cite{lapidoth1998poisson} shows that the mutual information function over $\mu_1$ and $\mu_2$ is schur-concave for continuous time Poisson channel, while does not hold for non-perfect receiver. Section~\ref{subsect.majorization} demonstrates that $I_{X_1^2;Z}(\mu_1,\mu_2)$ is schur-concave as $A<A_{th}$, and $I_{X_1^2;Z}(\mu_1,\mu_2)$ is schur-concave and schur-convex for $\mathcal{C}^{+}$ and $\mathcal{C}^{-}$, respectively, for $A\geq A_{th}$. Furthermore, we have that $I_{X_1^2;Z}(\mu_1,\mu_2)$ is schur-concave for any fixed peak power $A$ as $\tau\to0$, as shown in Lemma~\ref{lemma.sisomacsymLemAth}.
\begin{lemma}\label{lemma.sisomacsymLemAth}
	For dead time $\tau\to0$, we have $\lim\limits_{\tau\to0}A_{th}=+\infty$, i.e., $I_{X_1^2;Z}(\mu_1,\mu_2)$ is schur-concave for any bounded peak power $A$.
	\begin{proof}
		Please refer to Appendix~\ref{appen.sisomacsymLemAth}.
	\end{proof}
\end{lemma} 

The same behavior of mutual information function between continuous Poisson channel and non-perfect receiver with small enough dead time, schur-concavity over any peak power $A$ and background radiation $\Lambda_0$, aligns with the intuition since small enough dead time would not cause any photon-counting loss.

\section{Sum-Rate MISO Capacity for Two Users}\label{sect.misomac}
We extend the analysis to the case when the user $m$ is equipped with $J_m$ (more than one) transmitters.
\subsection{Sum-Rate MISO-MAC Capacity Analysis}
The sum-rate MISO-MAC capacity is defined as $C_{MISO-MAC}\dff\max\limits_{\Lambda^{T_s}_{mj}\in[0,A]}\frac{1}{T_s}I(\Lambda^{T_s}_{[MJ]};Z)$.
Similar to Section \ref{sect.sisomac}, the input waveform signal of each transmitter is piece-wise constant waveforms with two
levels $\{0,A_{mj}\}$ for the $j^{th}$ transmitter of the $m^{th}$ user. Nevertheless, it is still needed to be investigated how the $J_m$ transmitters jointly work, which is addressed in the following result.
\begin{theorem}\label{theo.misomacall}
	For $\tau\leq\frac{\ln2}{\sum_{m=1}^{2}\sum_{j=1}^{J}A_{mj}+\Lambda_0}$, the optimal solution is that all transmitters must have the same duty cycle and must be on or off simultaneously. 
\end{theorem} 

Theorem \ref{theo.misomacall} implies the equivalence between MISO-MAC and SISO-MAC, i.e., the sum-rate MISO-MAC capacity with peak constraints $(A_{1[J]},A_{2[J]})$ is equivalent to the SISO-MAC capacity with peak power constraint $(\sum_{j=1}^{J}A_{1j},\sum_{j=1}^{J}A_{1j})$. Therefore, further detailed investigations on the sum-rate MISO-MAC capacity is similar to that in Section \ref{sect.sisomac}, and thus omitted here.

The proof of Theorem \ref{theo.misomacall} consists of the following two major steps.

In \textbf{step 1}, given duty cycle $\mu_{[2J]}$, we show how each transmitter jointly work.
\begin{proposition}\label{prop.misomacwork}
	For $\tau\leq\frac{\ln2}{\sum_{m=1}^{2}\sum_{j=1}^{J}A_{mj}+\Lambda_0}$, the optimal condition is that if the transmitter with a smaller duty cycle is on then all transmitters with a larger duty cycle must also be on. 
	\begin{proof}
		Please refer to Appendix \ref{appen.misomacwork}.
	\end{proof}
\end{proposition}

This proposition shows that the capacity-achieving transmitted signals through $J_{m}$ transmitter are correlated but i.i.d. in each time interval for user $m$, $m=1,2$.The possible PMF value can be reduced from $2^{J_1}+2^{J_2}$ to $J_1+J_2+2$. Similarly, for a given duty cycle, there could be infinite number of possible joint distribution. The main idea is to show that, if the transmitter
with lower duty cycle is on, then the transmitter with higher duty cycle must be also on to achieve optimality. In addition, the objective function is concave and the optimization is performed over a convex compact set, such that the optimal solution clearly exists.

In \textbf{step 2}, we show the following proposition that characterizes the optimal duty cycle.
\begin{proposition}\label{prop.misomacwork2}
	For the optimal solution, all transmitters of user $m$ must have the same duty cycle, and according to Proposition \ref{prop.misomacwork} they must be on and off simultaneously.
	\begin{proof}
		Please refer to Appendix \ref{appen.misomacwork2}.
	\end{proof}
\end{proposition}

This proposition shows that for the optimal solution, the transmitters of each user must have the same duty cycle (i.e., $\mu_{m1}=\cdots=\mu_{mJ_m}$) and must be aligned. Hence, the dimension of the optimization
problem can be reduced from $J_1+J_2$ to $2$. The main idea is to show that all transmitters are simultaneously on and off. Hence, from the
receiver perspective, two users with multiple transmitters can be viewed as two users each with a single transmitter with peak power constraint $(\sum_{j=1}^{J_1}A_{1j},\sum_{j=1}^{J_2}A_{2j})$.
\section{Numerical Results}\label{sect.numresult}
In this section, we provide numerical examples to illustrate
results obtained in this paper. As shown in the paper,
the MISO-MAC Poisson capacity can be converted to that of SISO-MAC
. Hence, in the following, we provide only example
related to the SISO-MAC case.

Figure~\ref{fig.MACfgnonf} and Figure~\ref{fig.MACfgone} show the case of no intersection and one intersection in $0\leq\mu_1\leq1$ and $0\leq\mu_2\leq1$, respectively. The dead time is set to $0.02$; The peak power $(A_1,A_2)$ of transmitters $1$ and $2$ are set to $(1,20)$ and $(10,12)$, respectively, in Figures~\ref{fig.MACfgnonf} and \ref{fig.MACfgone} and satisfies the condition $\tau\leq\frac{\ln2}{A_1+A_2+\Lambda_0}$. Lemma~\ref{lemma.sisomacconvexg} implies at most two intersection points between function $f_{MAC}(\cdot)$ and $g_{MAC}(\cdot)$, while can not find the case of two intersection points by brute-force search. 
Figure~\ref{fig.MACmucomb} illustrates the optimal $\mu_1$ and $\mu_2$ against peak power of user 2 $A_2$ for different dead time $\tau$ given $A_1=12.5$. $\tau=0$ represents continuous Poisson channel. It is seen that $\tau\leq\frac{\ln2}{A_1+A_2+\Lambda_0}$ is satisfied and the optimal $\mu_1$ and $\mu_2$ close to that of continuous Poisson channel as $\tau\to0$ and the optimal $\mu_1^{*}=\mu_2^{*}$ as $A_1=A_2$ for any dead time $\tau$, aligned with the result of Section~\ref{sect.sisomac2}. Figure~\ref{fig.MACcapa} shows the MAC Poisson capacity with respect to peak power $A_2$ for different dead time and it is seen that the optimal MAC Poisson capacity with non-perfect receiver approaches that of continuous Poisson channel as $\tau\to0$, aligned with Theorem~\ref{theo.macasymp}.
Figure~\ref{fig.macregion} shows the optimal transmission strategy region of $A_1$ and $A_2$. ``Black'', ``red'', and ``Blue'' regions represents the optimal transmission strategy region of only active user $2$, both two active users and only active user $1$, respectively. It is seen that the boundary of these three regions are almost two lines through the origin with different slope, and the optimal transmission strategy are only user $2$-active, both two-user-active and only user $1$-active for the case of $A_1\ll A_2$, $A_1= A_2$, $A_1\gg A_2$, respectively, aligned with Section~\ref{subsect.stragety} and Section~\ref{sect.sisomac2}.

\section{Conclusion}\label{sect.conclusion}
In this paper, we have characterized the two-user asymmetric sum-rate Poisson capacity for both SISO and MISO cases. We demonstrate the equivalence of these two cases under certain condition. For both two cases, the optimal input signal of each transmitter and user must be two-level piece-wise constant and there are three possible transmission strategies, including only one active user and both active users. We provide the sufficient condition of these three strategies. In addition, we investigate the two-user symmetric sum-rate Poisson capacity based on above result and majorization method, both demonstrating that the optimal duty cycle must be the same and unique, and the majorization method maybe can be extend to multiple users case. 
\begin{figure}
	\setlength{\abovecaptionskip}{-0.1cm}
	\centering
	{\includegraphics[angle=0, width=0.8\textwidth]{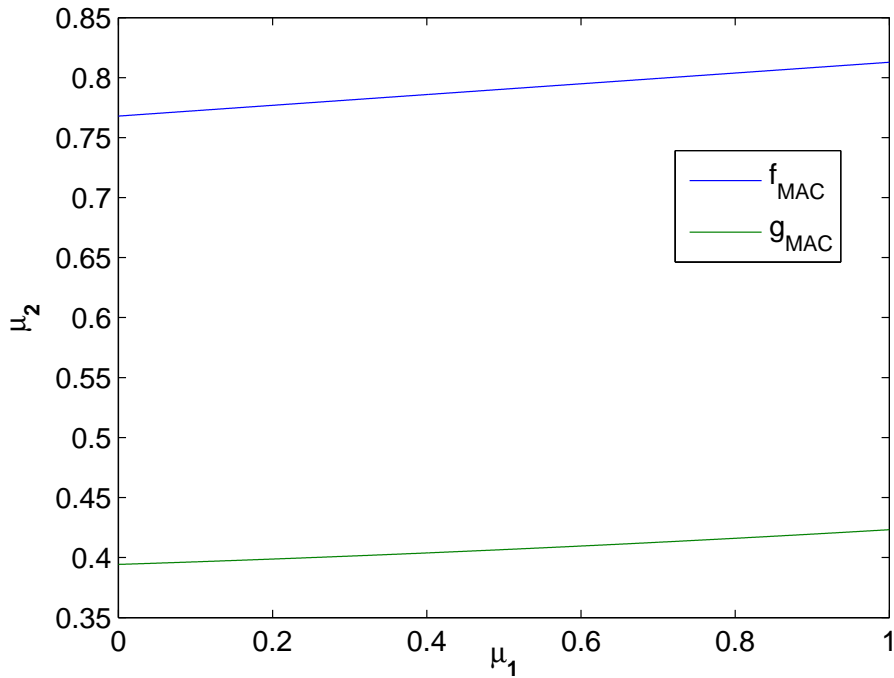}}
	\caption{$f_{MAC}(\mu_1)$ and $g_{MAC}(\mu_1)$ have no intersection in $0\leq\mu_1\leq1$ and $0\leq\mu_2\leq1$.}
	\label{fig.MACfgnonf}	
\end{figure}
\begin{figure}
	\centering
	{\includegraphics[angle=0, width=0.8\textwidth]{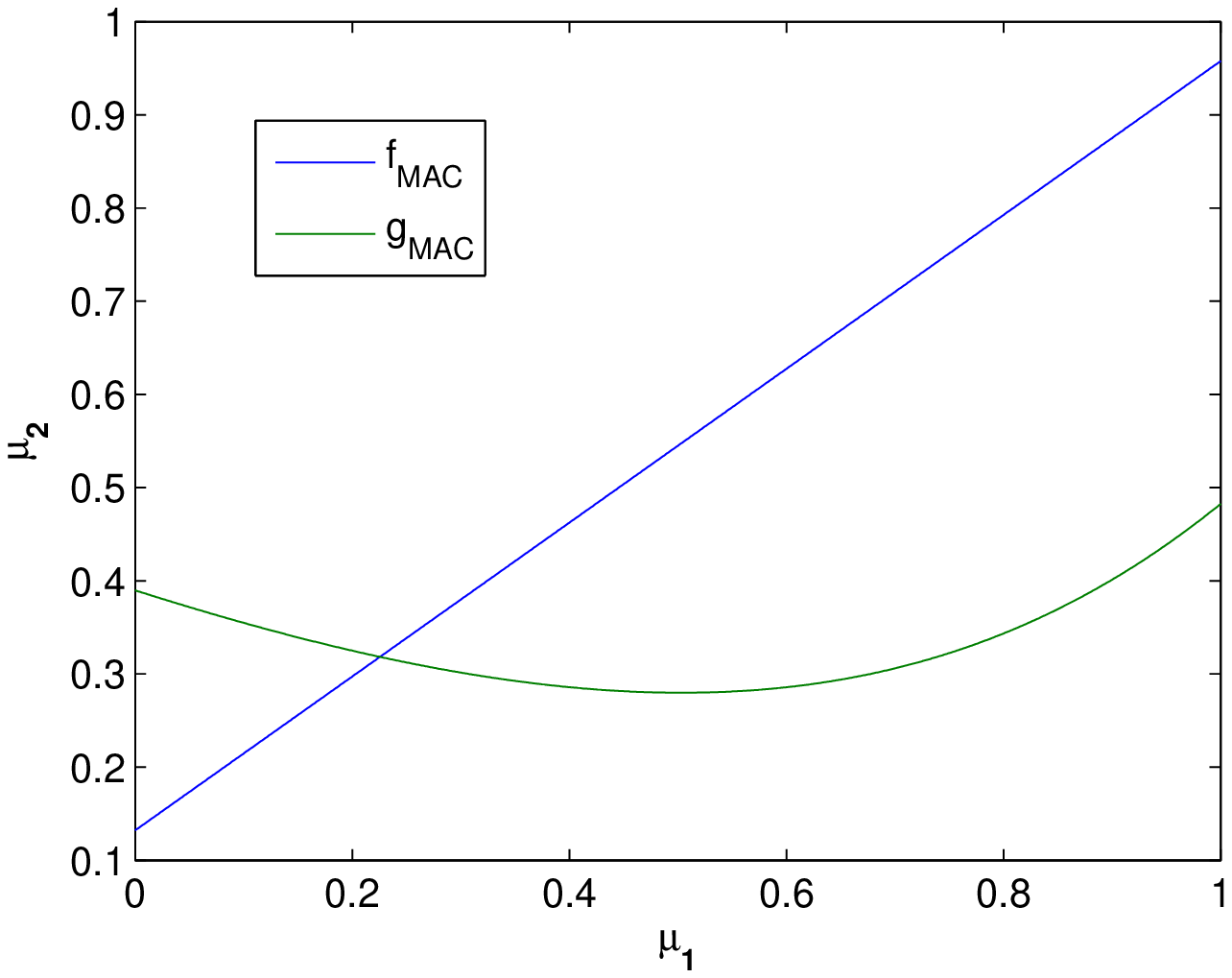}}
	\caption{$f_{MAC}(\mu_1)$ and $g_{MAC}(\mu_1)$ have one intersection in $0\leq\mu_1\leq1$ and $0\leq\mu_2\leq1$.}
	\label{fig.MACfgone}	
\end{figure}
\begin{figure}
	\centering
	{\includegraphics[angle=0, width=0.8\textwidth]{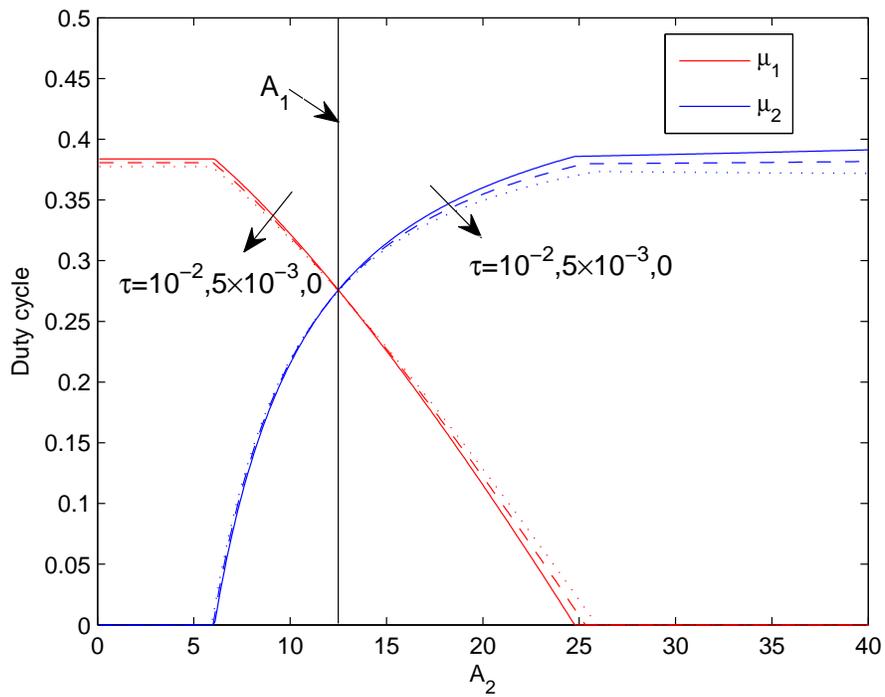}}
	\caption{The optimal $\mu_1$ and $\mu_2$ versus peak power of user 2 $A_2$ for different dead time $\tau$.}
	\label{fig.MACmucomb}	
\end{figure}
\begin{figure}
	\centering
	{\includegraphics[angle=0, width=0.8\textwidth]{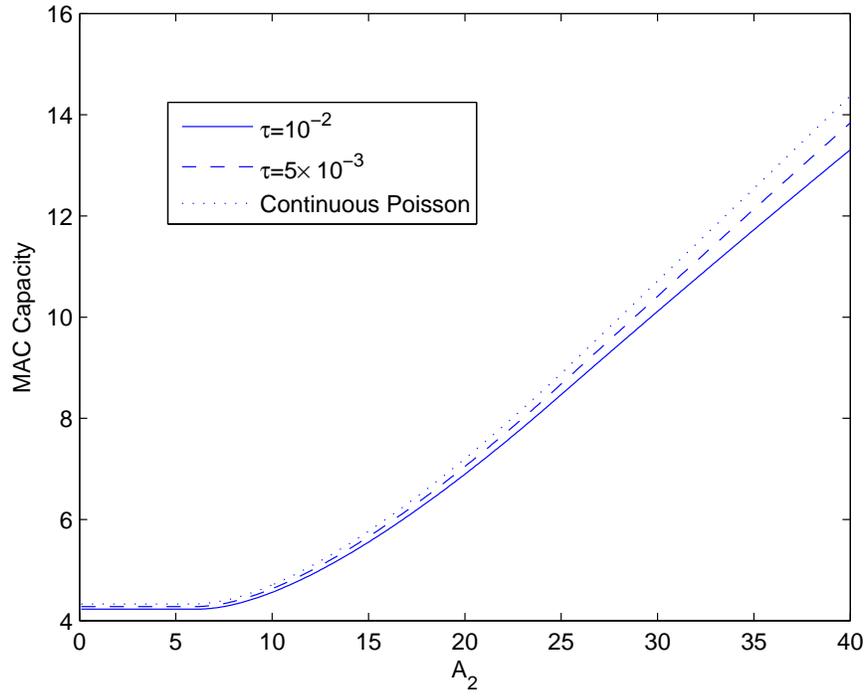}}
	\caption{The MAC Poisson capacity versus peak power $A_2$ for different dead time.}
	\label{fig.MACcapa}
\end{figure}
\begin{figure}
		\centering
		{\includegraphics[angle=0, width=0.8\textwidth]{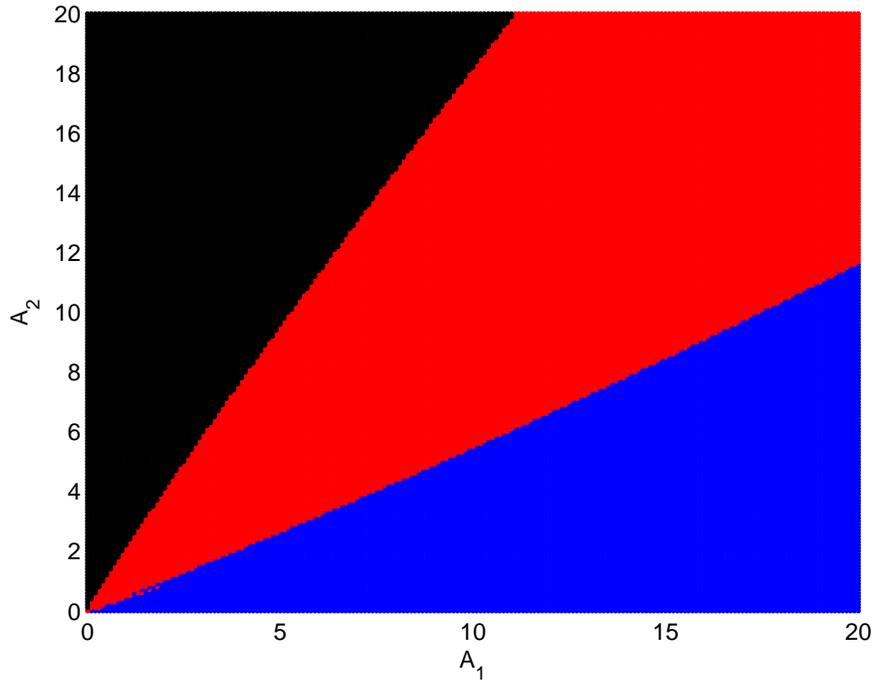}}
		\caption{Optimal transmission strategy for different peak power $A_1$ and $A_2$.}
		\label{fig.macregion}	
\end{figure}


\appendices
\renewcommand{\baselinestretch}{1.4}
\section{The proof of main results on MISO-MAC Capacity}
\subsection{Proof of Theorem \ref{theo.sisomacbi}}\label{appen.sisomacbi}
\textbf{Converse part}: Note that $\Lambda^{T_s}_{[M]}\rightarrow X_{[M]}\rightarrow Z$ forms a Markov chain, according to DPI, we have $I(\Lambda^{T_s}_{[M]};Z)\leq I(X_{[M]};Z)$.
The mutual information $I(X,Z)$ is as follow,
\be 
I(X,Z)=h_b(\hat{p})-\int h_b(p(\sum_{m=1}^{M}X_m+\Lambda_0))\mathrm{d}\mu(X_{[M]}),
\ee 
where $\hat{p}=\mathbb{E}[p(\sum_{m=1}^{M}X_m+\Lambda_0)]$. Define $S_1=p(X_1+\Lambda_0)$, note that the mapping $X\rightarrow S$ is a one-to-one mapping and $p(x_1+x_2)=p(x_2)+(1-p(x_2))p(x_1)$, hence we have 
\be 
I(X_{[M]};Z)=I(X_{2}^M,S_1;Z)=h_b(\hat{p})+\mathbb{E}_{X_{2}^M}\Big[\mathbb{E}_{S_1}[-h_b\big(p(X_{2}^M)+(1-p(X_{2}^M))S_1\big)]\Big],
\ee
and the following equation holds,
\be
&&\max\limits_{\mu(X_{[M]})}I(X_{[M]},Z)=\max\limits_{\mu(,X_{2}^M,S_1)}I(X_{2}^M,S_1;Z)\nonumber\\&&=\max\limits_{p(\Lambda_0)\leq\hat{p}\leq p(A+\Lambda_0)}\hat{p}+\max\limits_{\mu(X_{2}^M)}\mathbb{E}_{X_{2}^M}\Big[\max\limits_{\mu(S_1)\in\mathcal{S}_{M}(\hat{p},\mu(X_2^M))}\mathbb{E}_{S_1}[-h_b\big(p(X_{2}^M)+(1-p(X_{2}^M))S_1\big)]\Big],
\ee 
where 
\be
\mathcal{S}_M(\hat{p},\mu(X_2^M))=\Big\{\mu(S_1):\mathbb{E}[S]=\frac{\hat{p}-\mathbb{E}[p(\sum_{m=2}^{M}A_m)]}{1-\mathbb{E}[p(\sum_{m=2}^{M}A_m)]}\Big\},
\ee
where the inner optimization is performed over the class of distributions
of $S_1$ with a finite support $[0,A_1]$ and fixed conditional
mean set. Note that convex function $-h_b(\cdot)$ compounded linear function is still a convex function, the inner maximum is achieved if and only if $S_1$ is two-levels. Then we see that the optimal marginal PMF of $X_1$ is given by
\be 
\mathbb{P}(X_1=A_1)=\mathbb{P}\big(S_1=p(A_1+\Lambda_0)\big)\dff\mu_1.
\ee 
By symmetry, the optimal marginal PMF of $X_m$ is $\{0,A_m\}$-valued with $\mathbb{P}(X_m=A_m)\dff\mu_m$, where $m=2,\cdots,M$.

\textbf{Achievability part:} Let waveform $\Lambda_m^{T_s}$ in $[0,T_s]$ randomly selected from waveform set $\{0,A_m*(u(t)-u(t-T_s))\}$ with probability $\mu_m^{*}=\mathbb{P}\{\Lambda_m^{T_s}=A_m*(u(t)-u(t-T_s))\}$, where $u(t)$ denotes as a step function, then we have the upper bound in converse part is achievable.

\subsection{Proof of Lemma \ref{lemmal.sisomacnonconacve}}\label{appen.sisomacconcave}
\textbf{Non-convex optimized joint distribution set}:
The joint distribution of two independent variable is not closed under the linear weighted operation, i.e.,
\be 
\mu f_{X^{'}_{[2]}}(x_1,x_2)+(1-\mu)\mu f_{X^{''}_{[2]}}(x_1,x_2)&\neq&\int\mu f_{X^{'}_{[2]}}(x_1,x_2)+(1-\mu)\mu f_{X^{''}_{[2]}}(x_1,x_2)\mathrm{d}x_1\nonumber\\&&\cdot\int\mu f_{X^{'}_{[2]}}(x_1,x_2)+(1-\mu)\mu f_{X^{''}_{[2]}}(x_1,x_2)\mathrm{d}x_2,
\ee 
where $f_{X_{[2]}}(x_1,x_2)$ denotes as the joint distribution of $X_{[2]}$.

\textbf{Non-concavity of $I_{X_1^2;Z}(\mu_1,\mu_2)$}: Prove by contradiction.
Assume $I_{X_1^2;Z}(\mu_1,\mu_2)$ is concave, $\nabla^2 I$ needs to be negative
semi-definite. By calculating, we have
\be 
\frac{\partial I}{\partial \mu_1}&=&\ln\frac{1-\hat{p}}{\hat{p}}\big(\mu_2(p_1-p_2)+(1-\mu_2)(p_3-p_4)\big)\nonumber\\&&-\big(\mu_2(h_b(p_1)-h_b(p_2))+(1-\mu_2)(h_b(p_3)-h_b(p_4))\big),\\
\frac{\partial I^2}{\partial^2 \mu_1}&=&-\frac{\big(\mu_2(p_1-p_2)+(1-\mu_2)(p_3-p_4)\big)^2}{\hat{p}(1-\hat{p})}<0,\\
\frac{\partial I^2}{\partial^2 \mu_1\mu_2}&=&-\underbrace{\frac{1}{\hat{p}(1-\hat{p})}\big(\mu_1(p_1-p_3)+(1-\mu_1)(p_2-p_4)\big)\big(\mu_2(p_1-p_2)+(1-\mu_2)(p_3-p_4)\big)}_{I_{a}}\nonumber\\&&+\underbrace{\ln\frac{1-\hat{p}}{\hat{p}}(p_1-p_2-p_3+p_4)-\big(h_b(p_1)-h_b(p_2)-h_b(p_3)+h_b(p_4)\big)}_{I_{b}},\\
\frac{\partial I^2}{\partial^2 \mu_2}&=&-\frac{\big(\mu_1(p_1-p_3)+(1-\mu_1)(p_2-p_4)\big)^2}{\hat{p}(1-\hat{p})}<0,
\ee 
Thus $|\nabla^2 I|$ is given by
\be 
|\nabla^2 I|&=&\frac{\partial I^2}{\partial^2 \mu_1} \frac{\partial I^2}{\partial^2 \mu_2}-(\frac{\partial I^2}{\partial \mu_1\partial \mu_2})^2\nonumber\\&=&\Big[\ln\frac{1-\hat{p}}{\hat{p}}(p_1-p_2-p_3+p_4)-\big(h_b(p_1)-h_b(p_2)-h_b(p_3)+h_b(p_4)\big)-\frac{\partial I^2}{\partial^2 \mu_1\mu_2}\Big]^2\nonumber\\&&-(\frac{\partial I^2}{\partial^2 \mu_1\mu_2})^2=I_{b}(2I_{a}-I_{b}),
\ee 

Note that 
\be 
\lim\limits_{(\mu_1,\mu_2)\to\mathbf{0}}I_{b}(2I_{a}-I_{b})&=&(p_1-p_2-p_3+p_4)\Big(\ln\frac{1-p_4}{p_4}-\bar{G}(A_1,A_2)\Big)\Big(2\frac{(p_2-p_4)(p_3-p_4)}{p_4(1-p_4)}\nonumber\\&&-(p_1-p_2-p_3+p_4)\big(\ln\frac{1-p_4}{p_4}-\bar{G}(A_1,A_2)\big)\Big)\nonumber\\
&\overset{(a)}{=}&(p_1-p_2-p_3+p_4)^2\big(\ln\frac{1-p_4}{p_4}-\bar{G}(A_1,A_2)\big)\Big(\bar{G}(A_1,A_2)+\frac{2}{p_4}-\ln\frac{1-p_4}{p_4}\Big).
\ee 
where $\bar{G}(A_1,A_2)\dff\frac{h_b(p_1)-h_b(p_2)-h_b(p_3)+h_b(p_4)}{p_1-p_2-p_3+p_4}>0$, equality $(a)$ holds since $(p_2-p_4)(p_3-p_4)=-(p_2-p_4)(1-p_4)p(A_1)=-(1-p_4)(p_1-p_2-p_3+p_4)=$, $\bar{G}(A_1,A_2)+\frac{2}{p_4}-\ln\frac{1-p_4}{p_4}>0$ since $\frac{1}{p_4}>\frac{1}{p_4}>\frac{1-p_4}{p_4}$. Set $\tau=0.02$, $\Lambda_0=0.001$ and $A_1=A_2=10$, we have $\tau=0.02\leq\frac{\ln2}{A_1+A_2+\Lambda_0}\simeq0.0347$ and $\bar{G}(10,10)\simeq9.51<\ln\frac{1-p_4}{p_4}\simeq10.8198$, i,e., there exists certain $(A_1,A_2,\Lambda_0)$ such that $\lim\limits_{(\mu_1,\mu_2)\to\mathbf{0}}I_{b}(2I_{a}-I_{b})>0$.

Thus, there exists $\epsilon>0$ so that for $(\mu_1,\mu_2)\in \mathcal{C}_{\epsilon}=\{(\mu_1,\mu_2):\sqrt{\mu_1^2+\mu_2^2}\leq\epsilon\}$,  we have $|\nabla^2 I|>0$, contradicted with assumption of negative
semi-definite $\nabla^2 I$. 

\subsection{Proof of Lemma \ref{lemma.pnuvw}}\label{appen.pnuvw}
According to Lemma \ref{appenAu.convex} and $h_b(\cdot)$ is concave, we have
\be 
\frac{h_b(p_1)-h_b(p_2)}{p_1-p_2}>\frac{h_b(p_1)-h_b(p_4)}{p_1-p_4}>\frac{h_b(p_3)-h_b(p_4)}{p_3-p_4},\\
\frac{h_b(p_2)-h_b(p_3)}{p_2-p_3}>\frac{h_b(p_2)-h_b(p_4)}{p_2-p_4}.
\ee
By calculating, we can have 
\be 
U=(p_1-p_2)(p_3-p_4)[\frac{h_b(p_1)-h_b(p_2)}{p_1-p_2}-\frac{h_b(p_3)-h_b(p_4)}{p_3-p_4}]>0,
\ee 
similarly, we can have $V>0$.
As for $W$, if $A_2>A_1$, we have $p_2>p_3$ and
\be 
W&=&p_2p_3\frac{h_b(p_2)-h_b(p_3)}{p_2-p_3}-p_2p_4\frac{h_b(p_2)-h_b(p_4)}{p_2-p_4}+p_3p_4\frac{h_b(p_2)-h_b(p_4)}{p_2-p_4}\nonumber\\
&>&p_2p_4[\frac{h_b(p_2)-h_b(p_3)}{p_2-p_3}-\frac{h_b(p_2)-h_b(p_4)}{p_2-p_4}]+p_3p_4\frac{h_b(p_2)-h_b(p_4)}{p_2-p_4}>0
\ee 
Similarly, we have $W\lesseqqgtr0$ if and only if $A_1\gtreqqless A_2$.
\subsection{Proof of Convexity of $g_{MAC}(\mu_1)$}\label{appen.convexgmac}
Define $d_0\dff p_2-p_4>0$, $d_1\dff p_1-p_2-p_3+p_4<0$, $d_2\dff h_b(p_1)-h_b(p_2)-h_b(p_3)+h_b(p_4)<0$, and $d_3\dff h_b(p_2)-h_b(p_4)$. Note that $\frac{h_b(p_2)-h_b(p_4)}{p_2-p_4}>\frac{h_b(p_1)-h_b(p_3)}{p_1-p_3}$ and $p_2-p_4>p_1-p_3>0$, we have
\be 
\frac{d_2}{d_1}=\frac{[h_b(p_2)-h_b(p_4)]-[h_b(p_1)-h_b(p_3)]}{(p_2-p_4)-(p_1-p_3)}>\frac{h_b(p_2)-h_b(p_4)}{p_2-p_4}=\frac{d_3}{d_0},
\ee 
which implies $d_2d_0-d_1d_3=d_0d_1(\frac{d_2}{d_1}-\frac{d_3}{d_0})<0$. Note that $a_M=\exp(\frac{\mu_1d_2+d_3}{\mu_1d_1+d_0})$ and $d_0+d_1>0$, we have
\be 
a_M^{'}&=&\exp(\frac{\mu_1d_2+d_3}{\mu_1d_1+d_0})\frac{d_2d_0-d_1d_3}{\mu_1d_1+d_0}<0,\label{eq.firstdaM}\\
a_M^{''}&=&a_M^{'}(\mu_1d_1+d_0)^{-2}[(d_2d_0-d_1d_3)-2d_1(\mu_1d_1+d_0)],
\ee 
Note that $\mu_1(p_1-p_3)+(1-\mu_1)(p_2-p_4)=\big(1-(1-p(A_1))\mu_1\big)(p_2-p_4)$, after rearrangement of $g_{MAC}(\mu_1)$, we have
\be\label{eq.gmac}
g_{MAC}(\mu_1)=\frac{1}{F(\mu_1)}-\frac{\mu_1(p_3-p_4)+p_4}{\big(1-(1-p(A_1))\mu_1\big)(p_2-p_4)},
\ee 
where $F(\mu_1)\dff (a_M+1)\big(1-(1-p(A_1))\mu_1\big)(p_2-p_4)>0$. It is easy to check 
\be 
F^{'}(\mu_1)&=&(p_2-p_4)\{a_M^{'}\big(1-(1-p(A_1))\mu_1\big)-(1-p(A_1))(a_M+1)\}<0,\\
F^{''}(\mu_1)&=&a_M^{'}(\mu_1d_1+d_0)^{-2}\big\{[(d_2d_0-d_1d_3)-2d_1(\mu_1d_1+d_0)]\big(1-\mu_1(1-p(A_1))\big)\nonumber\\&&-2(1-p(A_1))(\mu_1d_1+d_0)^2\}\dff a_M^{'}(\mu_1d_1+d_0)^{-2}l(\mu_1)
\ee 
It is easy to check that $l(\mu_1)$ is a linear function. Note that $d_1=-d_0(1-p(A_1))$, we have
\be 
l(1)&=&[(d_2d_0-d_1d_3)-2d_1(d_1+d_0)]p(A_1)-2(1-p(A_1))(d_1+d_0)^2\nonumber\\&=&(d_2d_0-d_1d_3)p(A_1)<0,\\
l(0)&=&(d_2d_0-d_1d_3-2d_1d_0)-2(1-p(A_1))d_0^2=d_2d_0-d_1d_3<0,
\ee 
therefore, we have $l(\mu_1)<0$ and $F^{''}(\mu_1)<0$ for $\mu_1\in[0,1]$ and $[\frac{1}{F(\mu_1)}]^{''}=\frac{2}{F^3(\mu_1)}[F^{'}(\mu_1)]^2-\frac{1}{F^2(\mu_1)}F^{''}(\mu_1)>0$.

For the last fraction in Equation (\ref{eq.gmac}), we have its derivation as follows,
\be 
[-\frac{\mu_1(p_3-p_4)+p_4}{\big(1-(1-p(A_1))\mu_1\big)(p_2-p_4)}]^{''}&=&\frac{-2(1-p(A_1))[(p_3-p_4)+p_4(1-p(A_1))]}{\big(1-(1-p(A_1))\mu_1\big)^3(p_2-p_4)}\nonumber\\&=&\frac{-2(a_M+1)^3(p_2-p_4)^2(1-p(A_1))[p_3-p_4p(A_1)]}{F^3(\mu_1)},\nonumber
\ee 
Based on Equation (\ref{eq.gmac}), $g_{MAC}^{''}$ is given by
\be 
g_{MAC}^{''}&=&[\frac{1}{F(\mu_1)}]^{''}+[-\frac{\mu_1(p_3-p_4)+p_4}{\big(1-(1-p(A_1))\mu_1\big)(p_2-p_4)}]^{''}\nonumber\\
&\overset{(a)}{>}&\frac{2}{F^3(\mu_1)}+[F^{'}(\mu_1)]^2\frac{-2(a_M+1)^3(p_2-p_4)^2(1-p(A_1))[p_3-p_4p(A_1)]}{F^3(\mu_1)}\nonumber\\
&=&\frac{2(p_2-p_4)^2}{F^3(\mu_1)}\Big\{\Big[a_M^{'}\big(1-(1-p(A_1))\mu_1\big)-(1-p(A_1))(a_M+1)\Big]^2\nonumber\\&&-(a_M+1)^3(1-p(A_1))[p_3-p_4p(A_1)]\Big\}\nonumber\\
&\overset{(b)}{>}&\frac{2(p_2-p_4)^2(a_M+1)^2(1-p(A_1))}{F^3(\mu_1)}\Big\{(1-p(A_1))-(a_M+1)[p_3-p_4p(A_1)]\Big\}\nonumber\\
&\overset{(c)}{>}&\frac{2(p_2-p_4)^2(a_M+1)^2(1-p(A_1))}{F^3(\mu_1)}\Big\{(1-p(A_1))\nonumber\\&&-\Big(\exp\big(\frac{h_b(p_2)-h_b(p_4)}{p_2-p_4}\big)+1\Big)[p_3-p_4p(A_1)]\Big\}\nonumber\\
&\overset{(d)}{>}&\frac{2(p_2-p_4)^2(a_M+1)^2(1-p(A_1))}{F^3(\mu_1)}\Big\{(1-p(A_1))-(p_2+1)p_3\Big\}\nonumber\\
&\overset{(e)}{>}&\frac{2(p_2-p_4)^2(a_M+1)^2(1-p(A_1))}{F^3(\mu_1)}\big[(1-p(A_1)-p_3)p_2+(1-p_3)p(A_1)\big]>0
\ee 
where $(a)$ holds by dropping out positive terms $-\frac{1}{F^2(\mu_1)}F^{''}(\mu_1)$; $(b)$ holds since the term $\Big[a_M^{'}\big(1-(1-p(A_1))\mu_1\big)<0$ based on Equation (\ref{eq.firstdaM}), and $(1-p(A_1))(a_M+1)>0$; $(c)$ holds since $a_M\leq a_M(0)=\exp\big(\frac{h_b(p_2)-h_b(p_4)}{p_2-p_4}\big)$ based on Equation (\ref{eq.firstdaM}); $(d)$ holds since $\exp\big(\frac{h_b(p_2)-h_b(p_4)}{p_2-p_4}\big)>\exp\big(h_b^{'}(p_2)\big)=p_2$, and $(e)$ holds since $\tau\leq\frac{\ln2}{A_1+A_2+\Lambda_0}$, $p_1\leq\frac{1}{2}$, $p_1-p(A_1)=(1-p(A_1))p_2$ and $p_1-p_3=(1-p_3)p(A_1)$.
Thus, we have function $g_{MAC}(\mu_1)$ is a strictly convex function as $\tau\leq\frac{\ln2}{A_1+A_2+\Lambda_0}$.

\subsection{Proof of Lemma \ref{lemma.sisomacstra1}}\label{appen.sisomacstra1}
Note that $\lim\limits_{A_2\to\infty}p_1=\lim\limits_{A_2\to\infty}p_2=1$, and $\lim\limits_{A_2\to\infty}h_b(p_1)=\lim\limits_{A_2\to\infty}h_b(p_2)=0$, we have $\lim\limits_{A_2\to\infty}U=0$, and
\be 
\lim\limits_{A_2\to\infty}V=-\lim\limits_{A_2\to\infty}W=-h_b(p_4)+h_b(p_3)+p_3h_b(p_4)-p_4h_b(p_3),
\ee 
thus $\lim\limits_{A_2\to\infty}f_{MAC}(\mu_1)\equiv1$.
As for $\lim\limits_{A_2\to\infty}g_{MAC}$, it is straightforward that $(1+a_{MI})^{-1}<1$, thus we have $\lim\limits_{A_2\to\infty}g_{MAC}<1$ for any $\mu_1\in[0,1]$.
\subsection{Proof of Proposition \ref{prop.sisomacnec}}\label{appen.sisomacnec}
The main proof is based on continuity of $I_{X_1^2;Z}(\mu_1,\mu_2)$. When $\frac{\partial I}{\partial \mu_2}|_{(\tilde{\mu}_1,0)}>0$, for any $\epsilon$, there exists $0\mu_2<\epsilon$ so that $I_{X_1^2;Z}(\tilde{\mu}_1,0)<I_{X_1^2;Z}(\tilde{\mu}_1,\mu_2)$. Therefore, single user $1$ transmission is not optimal. Similarly, we have single user $2$ transmission is not optimal if $\frac{\partial I}{\partial \mu_2}|_{(0,\bar{\mu}_1)}>0$.

\subsection{Proof of Theorem \ref{theo.macasymp}}\label{appen.macasymp}
Similar to \cite{lai2017sum}, define $\phi(x)\dff x\ln x$ and $\alpha(x)\dff\frac{1}{x}\Big(e^{-1}(1+x)^{1+\frac{1}{x}}-1\Big)$. We first focus on the candidate optimal duty cycle for $3$ scenarios. For \textit{Scenario 2} of only active user $1$, since $h_b(x)=-x\ln(x)-(1-x)\ln(1-x)=x-x\ln(x)+o(x)$ and $\ln\frac{p(x)}{\tau}=\ln(x)+o(\tau)$, we have
\be\label{eq.prea1}
&&\frac{h_b(p_3)-h_b(p_4)}{p_3-p_4}+\ln\tau=\frac{A_1\tau-A_1\tau\frac{p_3}{\tau}+p_4\ln\frac{p_4}{\tau}+o(\tau)}{p_3-p_4} \\&=&1+\frac{\Lambda_0\ln\Lambda_0-(A_1+\Lambda_0)\ln(A_1+\Lambda_0)}{A_1}+o(1)=1-\ln(A_1+\Lambda_0)+s_1\ln\frac{s_1}{1+s_1}+o(1),\nonumber
\ee 
where $s_1=\frac{\Lambda_0}{A_1}$.
Thus, we have
\be \label{eq.optmu2}
&&\lim\limits_{\tau\to0}\tilde{\mu}_1=\lim\limits_{\tau\to0}\alpha_{\tau}(A_1,\Lambda_0)=\lim\limits_{\tau\to0}\frac{[1+\exp(\frac{h_b(p_3)-h_b(p_4)}{p_3-p_4})]^{-1}-p_4}{p_3-p_4}\nonumber\\
&=&\lim\limits_{\tau\to0}\frac{[\tau+\exp(\frac{h_b(p_3)-h_b(p_4)}{p_3-p_4}+\ln\tau)]^{-1}-\Lambda_0}{A_1}=\frac{1+s_1}{e}(\frac{s_1}{1+s_1})^{-s_1}-s_1=\alpha(s_1^{-1}).
\ee 
Equation (\ref{eq.optmu2}) implies that optimal duty cycle for \textit{Scenario 2} approaches that of continuous time as $\tau\to0$ \cite[Equation (22)]{lai2017sum}. Similar to \textit{Scenario 2}, the optimal duty cycle for \textit{Scenario 3} also approaches that of continuous time as $\tau\to0$.

For \textit{Scenario 1} of both active users, we focus on the asymptotic property of functions $g_{MAC}(\cdot)$ and $f_{MAC}(\cdot)$. Similar to Equation (\ref{eq.prea1}), we have
\be 
&&\frac{\mu_1\big(h_b(p_1)-h_b(p_3)\big)+(1-\mu_1)\big(h_b(p_2)-h_b(p_4)\big)}{\mu_1(p_1-p_3)+(1-\mu_1)(p_2-p_4)}+\ln\tau\\&=&\frac{\mu_1A_2\big(1+\phi(A_1+\Lambda_0)-\phi(A_1+A_2+\Lambda_0)\big)+(1-\mu_1)A_2\big(1+\phi(\Lambda_0)-\phi(A_2+\Lambda_0)\big)}{\mu_1A_2+(1-\mu_1)A_2}+o(1)\nonumber\\&=&
1+\mu_1A_2^{-1}\big(\phi(A_1+\Lambda_0)-\phi(A_1+A_2+\Lambda_0)\big)+(1-\mu_1)A_2^{-1}\big(\phi(\Lambda_0)-\phi(A_2+\Lambda_0)\big)+o(1),\nonumber
\ee 
and the function $g_{MAC}(\cdot)$ is given by
\be \label{eq.asymg1}
&&\lim\limits_{\tau\to0}g_{MAC}(\mu_1)=\lim\limits_{\tau\to0}\frac{(a_M\tau+\tau)^{-1}-[\mu_1(A_1+\Lambda_0)+(1-\mu_1)\Lambda_0]}{\mu_1A_2+(1-\mu_1)A_2}\nonumber\\&=&\frac{1}{A_2}\Big\{\exp\Big(-1-\mu_1A_2^{-1}\big(\phi(A_1+\Lambda_0)-\phi(A_1+A_2+\Lambda_0)\big)\nonumber\\&&-(1-\mu_1)A_2^{-1}\big(\phi(\Lambda_0)-\phi(A_2+\Lambda_0)\big)\Big)\Big\}-\frac{\mu_1A_1+\Lambda_0}{A_2}.
\ee 
which is aligned with \cite[Equation (19)]{lai2017sum}. For the function $f_{MAC}(\cdot)$, note that $(p_1-p_2)(p_3-p_4)=(1-p_2)(1-p_4)p^2(A_1)$ and $(p_1-p_3)(p_2-p_4)=(1-p_3)(1-p_4)p^2(A_2)$, we have
\be 
\lim\limits_{\tau\to0}\frac{(p_1-p_2)(p_3-p_4)}{(p_1-p_3)(p_2-p_4)}=\lim\limits_{\tau\to0}\frac{(1-p_2)p^2(A_1)}{(1-p_3)p^2(A_2)}=\frac{A_1^2}{A_2^2},
\ee 
Based on Equation (\ref{eq.prea1}) , we have 
\be\label{eq.asymf1} 
&&\lim\limits_{\tau\to0}\frac{U}{V}=\lim\limits_{\tau\to0}\frac{(p_1-p_2)\big(h_b(p_3)-h_b(p_4)\big)-(p_3-p_4)\big(h_b(p_1)-h_b(p_2)\big)}{(p_1-p_3)\big(h_b(p_2)-h_b(p_4)\big)-(p_2-p_4)\big(h_b(p_1)-h_b(p_3)\big)}\nonumber\\&=&
\frac{A_1^{-1}\Big(\phi(\Lambda_0)-\phi(A_2+\Lambda_0)-\phi(A_1+\Lambda_0)+\phi(A_1+A_2+\Lambda_0)\Big)}{A_2^{-1}\Big(\phi(\Lambda_0)-\phi(A_2+\Lambda_0)-\phi(A_1+\Lambda_0)+\phi(A_1+A_2+\Lambda_0)\Big)}\lim\limits_{\tau\to0}\frac{(p_1-p_2)(p_3-p_4)}{(p_1-p_3)(p_2-p_4)}\nonumber\\&=&\frac{A_1}{A_2},
\ee 
Note that $-W=p_2\big(h_b(p_3)-h_b(p_4)\big)-p_3\big(h_b(p_2)-h_b(p_4)\big)+p_4\big(h_b(p_2)-h_b(p_3)\big)=(p_2-p_4)\big(h_b(p_3)-h_b(p_4)\big)-(p_3-p_4)\big(h_b(p_2)-h_b(p_4)\big)$ and $\lim\limits_{\tau\to0}\frac{(p_2-p_4)(p_3-p_4)}{(p_1-p_3)(p_2-p_4)}=\lim\limits_{\tau\to0}\frac{p(A_2)p(A_1)}{p^2(A_2)}=\frac{A_1}{A_2}$, we have
\be \label{eq.asymf2}
&&\lim\limits_{\tau\to0}\frac{W}{V}=\lim\limits_{\tau\to0}\frac{(p_2-p_4)\big(h_b(p_3)-h_b(p_4)\big)-(p_3-p_4)\big(h_b(p_2)-h_b(p_4)\big)}{(p_1-p_3)\big(h_b(p_2)-h_b(p_4)\big)-(p_2-p_4)\big(h_b(p_1)-h_b(p_3)\big)}\nonumber\\&=&
\frac{A_1^{-1}\Big(\phi(\Lambda_0)-\phi(A_1+\Lambda_0)\Big)-A_2^{-1}\Big(\phi(\Lambda_0)-\phi(A_2+\Lambda_0)\Big)}{A_2^{-1}\Big(\phi(\Lambda_0)-\phi(A_2+\Lambda_0)-\phi(A_1+\Lambda_0)+\phi(A_1+A_2+\Lambda_0)\Big)}\lim\limits_{\tau\to0}\frac{(p_1-p_2)(p_3-p_4)}{(p_1-p_3)(p_2-p_4)}\nonumber\\&=&
\frac{\phi(\Lambda_0)-\phi(A_1+\Lambda_0)-\frac{A_1}{A_2}\Big(\phi(\Lambda_0)-\phi(A_2+\Lambda_0)\Big)}{\phi(\Lambda_0)-\phi(A_2+\Lambda_0)-\phi(A_1+\Lambda_0)+\phi(A_1+A_2+\Lambda_0)},
\ee 
which is aligned with \cite[Equation (16)]{lai2017sum}. Based on equations (\ref{eq.asymg1}), (\ref{eq.asymf1}) and (\ref{eq.asymf2}), all candidate optimal duty cycle approach to that of continuous time as $\tau\to0$.

For demonstrating the asymptotic property, we just need to show the asymptotic property of optimized objective function. According to L'Hospital's rule, we have
\be 
&&\lim\limits_{\tau\to0}\frac{I_{X_1^2;Z}(\mu_1,\mu_2)}{\tau}=\lim\limits_{\tau\to0}\frac{\partial I_{X_1^2;Z}(\mu_1,\mu_2)}{\partial \tau}\nonumber\\
&=&\lim\limits_{\tau\to0}\mu_1\mu_2(1-p_1)(A_1+A_2+\Lambda_0)\ln\frac{(1-\hat{p})p_1}{\hat{p}(1-p_1)}+(1-\mu_1)\mu_2(1-p_2)(A_2+\Lambda_0)\ln\frac{(1-\hat{p})p_2}{\hat{p}(1-p_2)}\nonumber\\
&&+\mu_1(1-\mu_2)(1-p_3)(A_1+\Lambda_0)\ln\frac{(1-\hat{p})p_3}{\hat{p}(1-p_3)}+(1-\mu_1)(1-\mu_2)(1-p_4)\Lambda_0\ln\frac{(1-\hat{p})p_4}{\hat{p}(1-p_4)}\nonumber\\
&=&\mu_1\mu_2\phi(A_1+A_2+\Lambda_0)+(1-\mu_1)\mu_2\phi(A_2+\Lambda_0)+\mu_1(1-\mu_2)\phi(A_1+\Lambda_0)\nonumber\\
&&+(1-\mu_1)(1-\mu_2)\phi(\Lambda_0)-\phi(\mu_1A_1+\mu_2A_2+\Lambda_0),
\ee 
which is aligned with \cite[Equation (8)]{lai2017sum}. Thus, the MAC Poisson capacity with non-perfect receiver approaches to that of continuous time.

\subsection{Proof of Lemma \ref{lemma.sisomacsymLem}}\label{appen.sisomacsymLem}
Since $\frac{\partial p_1}{\partial A}=2\tau(1-p_1)$ and $\frac{\partial p_2}{\partial A}=\tau(1-p_2)$, the derivative of $G(A)$ is given by
\be \label{eq.deriGA}
G^{'}(A)&=&\frac{2\tau}{(2p_2-p_1-p_4)^2}\Big\{\underbrace{[h_b^{'}(p_2)(1-p_2)-h_b^{'}(p_1)(1-p_1)]}_{>0}\underbrace{(2p_2-p_1-p_4)}_{>0}\nonumber\\&&-\underbrace{(p_1-p_2)}_{>0}\underbrace{\big(2h_b(p_2)-h_b(p_1)-h_b(p_4)\big)}_{>0}\Big\}\nonumber\\
&\overset{(a)}{=}&\frac{2\tau(p_1-p_2)}{(2p_2-p_1-p_4)}\Big\{\frac{(\ln p_1-\ln p_2)+\big(h_b(p_1)-h_b(p_2)\big)}{p_1-p_2}-\frac{2h_b(p_2)-h_b(p_1)-h_b(p_4)}{2p_2-p_1-p_4}\Big\}\nonumber\\
&\overset{(b)}{=}&\frac{2\tau(p_1-p_2)}{(2p_2-p_1-p_4)(p_2-p_4)p(A)\big(1-p(A)\big)}\Big\{[(\ln p_1-\ln p_2)+\big(h_b(p_1)-h_b(p_2)\big)]p(A)\nonumber\\&&-[2h_b(p_2)-h_b(p_1)-h_b(p_4)]\big(1-p(A)\big)\Big\}\nonumber\\
&=&\frac{2\tau(p_1-p_2)}{(2p_2-p_1-p_4)(p_2-p_4)p(A)\big(1-p(A)\big)}\Big\{[(\ln p_1-\ln p_2)]p(A)+\big(h_b(p_1)-h_b(p_2)\big)\nonumber\\&&-[h_b(p_2)-h_b(p_4)]\big(1-p(A)\big)\Big\}
\ee 
where $h_b^{'}(p_2)(1-p_2)-h_b^{'}(p_1)(1-p_1)<0$ holds since $\frac{\mathrm{d}}{\mathrm{d}x}(1-x)h_b^{'}(x)=-\frac{1}{x}-\ln\frac{1-x}{x}<0$, $2p_2-p_1-p_4>0$ holds due to concave function $p(\cdot)$ and $2h_b(p_2)-h_b(p_1)-h_b(p_4)>0$ holds according to Lemma~\ref{appenAu.hpconvex}; equality $(a)$ holds since $(1-x)h_b^{'}=-\ln x-h_b(x)$; inequality $(b)$ holds since $p_1-p_2=(p_2-p_4)\big(1-p(A)\big)$ and $2p_2-p_1-p_4=(p_2-p_4)p(A)$. Define $t=p(A)$, we have $p_1=p_4+(1-p_4)p(2A)=p_4+(1-p_4)(2t-t^2)$ and $p_2=p_4+(1-p_4)t$. For obtaining $(\ref{eq.deriGA})<0$, it is sufficient to show $\hat{G}(t,p_4)=[(\ln p_1-\ln p_2)]p(A)+\big(h_b(p_1)-h_b(p_2)\big)-[h_b(p_2)-h_b(p_4)]\big(1-p(A)\big)<0$, where $0\leq t\leq\frac{1}{2}$, and
\be 
\hat{G}(t)&\dff&t\ln\frac{p_4+(1-p_4)(2t-t^2)}{p_4+(1-p_4)t}+h_b\big(p_4+(1-p_4)(2t-t^2)\big)-h_b\big(p_4+(1-p_4)t\big)\nonumber\\&&-(1-t)\Big[h_b\big(p_4+(1-p_4)t\big)-h_b\big(p_4\big)\Big],
\ee 
It is easy to check that $\hat{G}(0,p_4)=0$ for any $p_4$ and $\hat{G}(t,0)=t\ln(2-t)+h_b\big(t(2-t)\big)-(2-t)h_b(t)$. Thus, we have
\be 
\frac{\partial \hat{G}(t,0)}{\partial t}&=&\ln(2-t)-\frac{t}{2-t}+h_b^{'}\big(t(2-t)\big)2(1-t)-(2-t)h_b^{'}(t)+h_b(t)\nonumber\\
&=&2(1-t)\big(\ln\frac{(1-t)^2}{t(2-t)}-\ln\frac{1-t}{t}\big)-\ln(1-t)+\ln(2-t)-\frac{t}{2-t}\nonumber\\
&=&(1-2t)\ln\frac{1-t}{2-t}-\frac{t}{2-t},\\
\frac{\partial^2 \hat{G}(t,0)}{\partial t^2}&=&-2\ln\frac{1-t}{2-t}-(1-2t)(\frac{1}{1-t}-\frac{1}{2-t})-\frac{2}{(2-t)^2}\nonumber\\
&=&2\ln(1+\frac{1}{1-t})-\frac{1-2t}{1-t}+\frac{t(2t-5)}{(2-t)^2}\\
&\overset{(c)}{\geq}&\frac{2}{2-t}-\frac{1-2t}{1-t}+\frac{t(2t-5)}{(2-t)^2}=\frac{t}{(1-t)(2-t)^2}>0.
\ee 
Thus we have $\frac{\partial \hat{G}(t,0)}{\partial t}\leq\hat{G}(\frac{1}{2},0)=-\frac{1}{3}<0$ and $\hat{G}(t,0)\leq\hat{G}(0,0)=0$. 

Similarly, for $p_4>0$, we have
\be 
\frac{\partial \hat{G}(t,p_4)}{\partial t}&=&\ln\frac{p_1}{p_2}+t(1-p_4)\underbrace{\big(\frac{2(1-t)}{p_1}-\frac{1}{p_2}\big)}_{<0}+2(1-t)(1-p_4)h_b^{'}(p_1)-(2-t)(1-p_4)h_b^{'}(p_2)\nonumber\\&&+\big(h_b(p_2)-h_b(p_4)\big)\nonumber\\
&<&2(1-t)(1-p_4)\ln\frac{(1-p_1)p_2}{p_1(1-p_2)}+\ln\frac{p_1}{p_2}-t(1-p_4)h_b^{'}(p_2)+\big(h_b(p_2)-h_b(p_4)\big)\nonumber\\
&\overset{(d)}{=}&2(1-t)(1-p_4)\ln\frac{(1-t)p_2}{p_1}+\ln\frac{p_1}{p_2(1-p_2)}+p_4h_b^{'}(p_2)-h_b(p_4)\nonumber\\
&=&[2(1-t)(1-p_4)-1]\ln\frac{(1-t)p_2}{p_1}+\underbrace{\ln\frac{1-t}{1-p_2}+p_4h_b^{'}(p_2)-h_b(p_4)}_{I_r<0}\nonumber\\
&\overset{(e)}{\leq}&[2(1-p_2)-1]\ln\frac{p_1-t}{p_1}<0
\ee 
where equality $(d)$ holds since $1-p_2=(1-p_1)(1-t)$, $t(1-p_4)=p_2-p_4$ and $p_2h_b^{'}(p_2)+h_b(p_2)=-\ln(1-p_2)$, $I_r<0$ since $\ln\frac{1-t}{1-p_2}+p_4h_b^{'}(p_2)-h_b(p_4)=\ln\frac{1}{1-p_4}+p_4h_b^{'}(p_2)-h_b(p_4)<\ln\frac{1}{1-p_4}+p_4h_b^{'}(p_4)-h_b(p_4)=0$, inequality $(e)$ holds since $1-p_2=(1-t)(1-p_4)$, $(1-t)p_2=p_1-t$ and $I_r<0$.

Thus, $G(A)$ decreases with peak power $A$. For peak power $A\to\infty$, we have $\lim\limits_{A\to\infty}G(A)=\lim\limits_{A\to\infty}\frac{2h_b(p_2)-h_b(p_1)-h_b(p_4)}{2p_2-p_1-p_4}=-\frac{h_b(p_0)}{1-p_0}$. For peak power $A\to0$, we have Taylor expansion on $p_1$ and $p_2$ as follows,
\be
p_1&=&p_4+(1-p_4)\tau\cdot 2A-\frac{1-p_4}{2}\tau^2\cdot (2A)^2+o(A^2),\\
p_2&=&p_4+(1-p_4)\tau\cdot A-\frac{1-p_4}{2}\tau^2\cdot A^2+o(A^2),
\ee 
thus we have $2p_2-p_1-p_4=(1-p_4)\tau^2A^2+o(A^2)$. Similarly, we have Taylor expansion on $2h_b(p_2)-h_b(p_1)-h_b(p_4)$ as follows,
\be 
2h_b(p_2)-h_b(p_1)-h_b(p_4)=-[h_b^{''}(p_4)(1-p_4)-h_b^{'}(p_4)](1-p_4)\tau^2A^2+o(A^2),
\ee 
and the limits $\lim\limits_{A\to0}G(A)$ is given by
\be 
\lim\limits_{A\to0}G(A)=\frac{-[h_b^{''}(p_4)(1-p_4)-h_b^{'}(p_4)](1-p_4)\tau^2}{(1-p_4)\tau^2}=\frac{1}{p_4}+\ln\frac{1-p_4}{p_4}>\ln\frac{1-p_4}{p_4}.
\ee 
Thus we have $G(A)\in(\ln(1-p_4)+\frac{p_4}{1-p_4}\ln p_4,\frac{1}{p_4}+\ln\frac{1-p_4}{p_4})$.

\subsection{Proof of Lemma \ref{lemma.sisomacsymLem2}}\label{appen.sisomacsymLem2}
Note that $\hat{p}\in[p_4,p_1]$ for $(\mu_1,\mu_2)\in[0,1]^2$, we have $\ln\frac{1-\hat{p}}{\hat{p}}\in[\ln\frac{1-p_1}{p_1},\ln\frac{1-p_4}{p_4}]$. Based on Lemma~\ref{appenAu.last} and $p_2-p_4>p_1-p_2$, we have $G(A)>\frac{h_b(p_2)-h_b(p_4)}{p_2-p_4}>\ln\frac{p_2}{p_2}>\ln\frac{p_1}{p_1}$. According to Lemma~\ref{lemma.sisomacsymLem} and $\lim\limits_{A\to0}G(A)=\frac{1}{p_4}+\ln\frac{1-p_4}{p_4}>\ln\frac{1-p_4}{p_4}$, there exists unique $A_{th}$ such that $\ln\frac{1-p_4}{p_4}=G(A)$ and the solution $\ln\frac{1-\hat{p}}{\hat{p}}=G(A)$ for $(\mu_1,\mu_2)\in[0,1]^2$ iff $A\geq A_{th}$.
\subsection{Proof of Theorem \ref{theo.sisomacsymBound}}\label{appen.sisomacsymBound}
It is easy to check that $G(A)\leq\ln\frac{1-p_4}{p_4}$ for $A\geq A_{th}$. Since $\ln\frac{1-\hat{p}}{\hat{p}}=G(A)$ and $\ln\frac{1-\hat{p}}{\hat{p}}$ decreases with $\hat{p}$, we have $\hat{p}=\frac{1}{1+\exp(G(A))}>p_4$. Note that continuous function $\hat{p}(\mu_1,\mu_2)$ increases with $\mu_1$ and $\mu_2$, thus, there exists differentiable function $f_B(\cdot)$ such that $\mathcal{C}=\{(\mu_1,\mu_2):\mu_1=f_B(\mu_2)\}$. 

According to Lemma~\ref{appenAu.last}, we have $G(A)>\frac{h_b(p_2)-h_b(p_4)}{p_2-p_4}>h_b^{'}(p_2)$ and $\frac{1}{1+\exp(G(A))}<p_2$. Note that the solution $\hat{p}(\mu_1,0)=\mu_1p_3+(1-\mu_1)p_4=\frac{1}{1+\exp(G(A))}\in(p_4,p_2)$ on $\mu_1$ exists, we have $0<f_B(0)<1$. For region $\mu_1\geq\mu_2$, we have
\be 
\frac{\partial \hat{p}}{\partial \mu_1}-\frac{\partial \hat{p}}{\partial \mu_2}&=&[\mu_2(p_1-p_2)+(1-\mu_2)(p_3-p_4)]-[\mu_1(p_1-p_3)+(1-\mu_1)(p_2-p_4)]\nonumber\\&=&(\mu_1-\mu_2)(2p_2-p_1-p_4)\geq0
\ee 
Take total differential on $\hat{p}(\mu_1,\mu_2)=\frac{1}{1+\exp(G(A))}$, we have 
\be\label{eq.firstdefB}
f_B^{'}(\mu_2)=\frac{\mathrm{d}\mu_1}{\mathrm{d}\mu_2}=-\frac{\frac{\partial \hat{p}}{\partial \mu_2}}{\frac{\partial \hat{p}}{\partial \mu_1}}\leq-1,
\ee
where the last inequality holds since $\frac{\partial \hat{p}}{\partial \mu_1}>\frac{\partial \hat{p}}{\partial \mu_2}>0$.

Since the cardinality of $|\mathcal{C}\cap\mathcal{L}_{\mu_s}|$ equals the number of intersect of $\mu_1=f_B(\mu_2)$ and $\mu_1+\mu_2=2\mu_s$ for $\mu_1\geq\mu_2$. Define $g_B(\mu_2)=f_B(\mu_2)-(2\mu_s-\mu_2)$, according to Equation (\ref{eq.firstdefB}), we have $g_B^{'}(\mu_2)=f_B^{'}(\mu_2)+1\leq0$ and the number of intersect of $\mu_1=f_B(\mu_2)$ and $\mu_1+\mu_2=2\mu_s$ for $\mu_1\geq\mu_2$ is at most $1$. Furthermore, we have $|\mathcal{C}\cap\mathcal{L}_{\mu_s}|=1$ iff
$f_B(0)\geq2\mu_s$ and $f_B(\mu_s)\leq\mu_s$. Define $f_B(\mu_s^{'})=\mu_s^{'}$ and $\mu_s^{*}=\frac{f_B(0)}{2}$, then we have $\mu_s^{*}=\frac{f_B(0)}{2}=\frac{\frac{1}{1+\exp(G(A))}-p_4}{2(p_2-p_4)}$ and $\mu_s^{*}<\frac{\tilde{\mu}_1}{2}$ since $G(A)>\frac{h_b(p_3)-h_b(p_4)}{p_3-p_4}$. In addition, for $\mu_s^{'}$, we have
\be\label{eq.musapos}
(2p_2-p_1-p_4)(\mu_s^{'})^2-2(p_2-p_4)\mu_s^{'}+\big(\frac{1}{1+\exp(G(A))}-p_4\big)=0
\ee 
Since $2p_2-p_1-p_4>0$, $p_2-p_4>0$ and $\frac{1}{1+\exp(G(A))}-p_4>0$, the two solutions on Equation (\ref{eq.musapos}) both are positive. Note that the summation of the two solutions equals to $\frac{2(p_2-p_4)}{2p_2-p_1-p_4}>2$, thus there exists unique feasible solution as follows,
\be 
\mu_s^{'}=\frac{(p_2-p_4)-\sqrt{(p_2-p_4)^2-(2p_2-p_1-p_4)[\frac{1}{1+\exp(G(A))}-p_4]}}{2p_2-p_1-p_4}.
\ee 
\subsection{Proof of Lemma \ref{lemma.sisomacsymLemAth}}\label{appen.sisomacsymLemAth}
For any fixed peak power $A$, we need to show that there exists $\tau>0$ such that $G(A)>\ln\frac{1-p_4}{p_4}$. The main clue is based on Taylor expansion of $\tau$.

Note that $p(x)=x\tau-\frac{1}{2}x^2\tau^2+o(\tau^2)$, we have
\be\label{eq.taylorp}
2p_2-p_1-p_4=\frac{1}{2}[(2A+\Lambda_0)^2+\Lambda_0^2-(A+\Lambda_0)^2]\tau^2+o(\tau^2)=A^2\tau^2+o(\tau^2).
\ee 
Since $h_b(x)=(1-\ln x)x+o(x)$, we have Taylor expansion of $h_b\big(p(x)\big)$ on $\tau$ as follows,
\be \label{eq.taylorhbp}
h_b\big(p(x)\big)&=&[1-\ln p(x)]p(x)+o\big(p(x)\big)=p(x)-p(x)\ln\tau+p(x)\ln(x+o(\tau))+o(\tau)\nonumber\\&=&p(x)-p(x)\ln\tau+[x\ln x] \tau+o(\tau).
\ee 
Similarly, we have 
\be\label{eq.taylorlnder}
\ln\frac{1-p_4}{p_4}=-\ln\tau-\ln\Lambda_0-\Lambda_0\tau+o(\tau),
\ee
Based on Equations (\ref{eq.taylorp}), (\ref{eq.taylorhbp}) and (\ref{eq.taylorlnder}), we have
\be 
G(A)-\ln\frac{1-p_4}{p_4}&=&\frac{2h_b(p_2)-h_b(p_1)-h_b(p_4)}{2p_2-p_1-p_4}-\ln\frac{1-p_4}{p_4}\nonumber\\
&=&(1-\ln\tau)+\frac{[2(A+\Lambda_0)\ln(A+\Lambda_0)-(2A+\Lambda_0)\ln(2A+\Lambda_0)-\Lambda_0\ln\Lambda_0]\tau}{A^2\tau^2}\nonumber\\
&&-[-\ln\tau-\ln\Lambda_0]+O(1)
\nonumber\\
&=&\frac{[2(A+\Lambda_0)\ln(A+\Lambda_0)-(2A+\Lambda_0)\ln(2A+\Lambda_0)-\Lambda_0\ln\Lambda_0]}{A^2}\frac{1}{\tau}+O(1).
\ee 
Since function $x\ln x$ is convex and $2(A+\Lambda_0)\ln(A+\Lambda_0)-(2A+\Lambda_0)\ln(2A+\Lambda_0)-\Lambda_0\ln\Lambda_0>0$, there exists $\tau>0$ such that $G(A)>\ln\frac{1-p_4}{p_4}$ for any fixed peak power $A$, i.e., $\lim\limits_{\tau\to0}A_{th}=+\infty$.

\subsection{Proof of Proposition \ref{prop.misomacwork}}\label{appen.misomacwork}
Define $b_{mj}(i_m)\in\{0,1\}$ as $j^{th}$ bit in the binary representation of '$i_m$' for user $m$, where $i_m=0,1,\cdots,2^{J_m}-1$, $i_m=\sum_{j_m=1}^{J_m}b_{mj_m}(i_m)2^{j-1}$. Define joint PMF $\mathbf{q}_{[M]}=(\mathbf{q}_1,\mathbf{q}_2)$, where $\mathbf{q}_m=\{q_{mi_m}\}_{i_m=0}^{2^{J_m}-1}$,  $q_{mi_m}\dff\mathbb{P}(X_{mj}=b_{mj_m}(i)A_{mj}),i_m=0,1,\cdots,2^{J_m}-1$. Then, $\mathbf{q}_{[M]}$ satisfies
\be\label{eq.macqconstr} 
q_{mi_m}&\geq&0, \quad i=0,1,\cdots,2^{J_m}-1;\nonumber\\
\sum_{i_m=0}^{2^{J_m}-1}q_{mi_m}&=&1;\\
\sum_{i_m=0}^{2^{J_m}-1}b_{mj_m}(i_m)q_{mi_m}&=&\mu_{mj_m},j_m=1,2,\cdots,J_m,\quad m=1,2.\nonumber
\ee 
Define $\hat{r}_{i_1i_2}=p(\sum_{m=1}^{2}\sum_{j_m=1}^{j_m}b_{mj_m}(i_m)A_{mi_m}+\Lambda_0)$. Noting that $\mu_{mj_m}=\sum_{i_m=0}^{2^J_m-1}b_{j_m}(i_m)q_{mj_m}$ for $j_m=1,2\cdots,J_m$, we have $C_{MISO-MAC}=\max\limits_{\mu_{[MJ]}\in[0,1]^{MJ}}\frac{1}{\tau}I_{MISO}(\mu_{[MJ]})$,
where
\be\label{eq.macjoint}
I_{MISO_MAC}(\mu_{[MJ]})&=&\max\limits_{\mathbf{q}_{M}}h_b\Big(\sum_{i_2=0}^{2^{J_2}-1}\sum_{i_1=0}^{2^{J_1}-1}q_{2i_2}q_{1i_1}\hat{r}_{i_1i_2}\Big)-\sum_{i_2=0}^{2^{J_2}-1}\sum_{i_1=0}^{2^{J_1}-1}q_{2i_2}q_{1i_1}h_b(\hat{r}_{i_1i_2})\nonumber\\
&=&\max\limits_{\mathbf{q}_{[M]}}h(\mathbf{q}_{[M]}).
\ee 
Noting that
\be 
\sum_{i_2=0}^{2^{J_2}-1}\sum_{i_1=0}^{2^{J_1}-1}q_{1i_1}q_{2i_2}\sum_{m=1}^{2}\sum_{j_m=1}^{J_m}b_{mj_m}(i_m)h_b\big(p(A_{mj_m})\big)&=&\sum_{m=1}^{2}\sum_{j_m=1}^{J_m}\sum_{i_2=0}^{2^{J_2}-1}\sum_{i_1=0}^{2^{J_1}-1}q_{1i_1}q_{2i_2}h_b\big(p(A_{mj_m})\big)\nonumber\\
&=&\sum_{m=1}^{2}\sum_{j_m=1}^{J_m}\mu_{mj_m}h_b\big(p(A_{mj_m})\big),
\ee 
we have
\be 
h(\mathbf{q_m})&=&h_b\Big(\sum_{i_2=0}^{2^{J_2}-1}\sum_{i_1=0}^{2^{J_1}-1}q_{2i_2}q_{1i_1}\hat{r}_{i_1i_2}\Big)-\sum_{i_2=0}^{2^{J_2}-1}\sum_{i_1=0}^{2^{J_1}-1}q_{1i_1}q_{2i_2}h_b(\hat{r}_{i_1i_2})\nonumber\\&&+\sum_{i_2=0}^{2^{J_2}-1}\sum_{i_1=0}^{2^{J_1}-1}q_{1i_1}q_{2i_2}\sum_{m=1}^{2}\sum_{j_m=1}^{J_m}b_{mj_m}(i_m)h_b\big(p(A_{mj_m})\big)-\sum_{m=1}^{2}\sum_{j_m=1}^{J_m}\mu_{mj_m}h_b\big(p(A_{mj_m})\big),\nonumber
\ee
\be
\frac{\partial h(\mathbf{q_{[M]}})}{\partial q_{mi_m}}&=&\sum_{i_{\bar{m}}=0}^{2^{J_{\bar{m}}}-1}\hat{r}_{i_mi_{\bar{m}}}\ln\frac{1-\sum_{i_2=0}^{2^{J_2}-1}\sum_{i_1=0}^{2^{J_1}-1}q_{2i_2}q_{1i_1}\hat{r}_{i_1i_2}}{\sum_{i_2=0}^{2^{J_2}-1}\sum_{i_1=0}^{2^{J_1}-1}q_{2i_2}q_{1i_1}\hat{r}_{i_1i_2}}-\sum_{i_{\bar{m}}=0}^{2^{J_{\bar{m}}}-1}q_{{\bar{m}}i_{\bar{m}}}h_b\big(\hat{r}_{i_mi_{\bar{m}}}\big)\nonumber\\&&+\sum_{i_{\bar{m}}=0}^{2^{J_{\bar{m}}}-1}q_{{\bar{m}}i_{\bar{m}}}\sum_{m=1}^{2}\sum_{j_m=1}^{J_m}b_{mj_m}(i_m)h_b\big(p(A_{mj_m})\big),\label{eq.deriveh}
\ee 
where $\bar{m}=3-m$. Let $I_{mi_m} =\{j_m\in\{1,2\cdots,J_m\} : b_{mj_m}(i_m)=1\}$ denotes the set of nonzero bit positions in the
binary representation of $i_m, i_m=0,1,\cdots,2^{J_m}-1$. Then, we will show that $I_{mi_m^{'}}\subseteq I_{mi_m}$ leads to
\be 
\frac{\partial h(\mathbf{q_{[M]}})}{\partial q_{mi_m^{'}}}\leq\frac{\partial h(\mathbf{q_{[M]}})}{\partial q_{mi_m}},\text{ for } i_m,i_m^{'}\in\{0,1,\cdots,2^{J_m}-1\}.
\ee 
For $\tau\leq\frac{\ln2}{\sum_{m=1}^{2}\sum_{j_m=1}^{J}A_{mj_m}+\Lambda_0}$, we have $\ln\frac{1-\sum_{i_2=0}^{2^{J_2}-1}\sum_{i_1=0}^{2^{J_1}-1}q_{2i_2}q_{1i_1}\hat{r}_{i_1i_2}}{\sum_{i_2=0}^{2^{J_2}-1}\sum_{i_1=0}^{2^{J_1}-1}q_{2i_2}q_{1i_1}\hat{r}_{i_1i_2}}>0$. According to lemma \ref{appenAu.hpconvex}, $h_b\big(p(x)\big)$ is concave. Based on Lemma \ref{appenAu.concavepro} and $h_b\big(p(0)\big)=0$, we have
\be 
\sum_{i_{\bar{m}}=0}^{2^{J_{\bar{m}}}-1}q_{{\bar{m}}i_{\bar{m}}}[h_b\big(\hat{r}_{i_mi_{\bar{m}}}\big)-h_b\big(\hat{r}_{i^{'}_mi_{\bar{m}}}\big)]&\leq& \sum_{i_{\bar{m}}=0}^{2^{J_{\bar{m}}}-1}q_{{\bar{m}}i_{\bar{m}}}h_b\big(p(\sum_{m=1}^{2}\sum_{j_m=1}^{J_m}[b_{mj_m}(i_m)-b_{mj_m}(i^{'}_m)]A_{mj_m})\big)\nonumber\\&\leq& \sum_{i_{\bar{m}}=0}^{2^{J_{\bar{m}}}-1}q_{{\bar{m}}i_{\bar{m}}}\sum_{m=1}^{2}\sum_{j_m=1}^{J_m}[b_{mj_m}(i_m)-b_{mj_m}(i^{'}_m)]h_b\big(p(A_{mj_m})\big).\nonumber
\ee 
According to equation (\ref{eq.deriveh}) and $r_i\geq r_{i^{'}}$, we have
\be 
\frac{\partial h(\mathbf{q_{[M]}})}{\partial q_{mi_m}}-\frac{\partial h(\mathbf{q_{[M]}})}{\partial q_{mi^{'}_m}}=\sum_{i_{\bar{m}}=0}^{2^{J_{\bar{m}}}-1}\underbrace{(\hat{r}_{i_mi_{\bar{m}}}-\hat{r}_{i^{'}_mi_{\bar{m}}})}_{\geq0}\underbrace{\ln\frac{1-\sum_{i_2=0}^{2^{J_2}-1}\sum_{i_1=0}^{2^{J_1}-1}q_{2i_2}q_{1i_1}\hat{r}_{i_1i_2}}{\sum_{i_2=0}^{2^{J_2}-1}\sum_{i_1=0}^{2^{J_1}-1}q_{2i_2}q_{1i_1}\hat{r}_{i_1i_2}}}_{\geq0}\\+\sum_{i_{\bar{m}}=0}^{2^{J_{\bar{m}}}-1}q_{{\bar{m}}i_{\bar{m}}}\underbrace{\Big\{\sum_{m=1}^{2}\sum_{j_m=1}^{J_m}b_{mj_m}(i_m)h_b\big(p(A_{mj_m})\big)-[h_b\big(\hat{r}_{i_mi_{\bar{m}}}\big)-h_b\big(\hat{r}_{i^{'}_mi_{\bar{m}}}\big)]\Big\}}_{\geq0}\geq0.
\ee 

Similar to \cite[Appendix B.2]{Chakraborty05}, property $\frac{\partial h(\mathbf{q_{[M]}})}{\partial q_{mi_m}}\geq\frac{\partial h(\mathbf{q_{[M]}})}{\partial q_{mi^{'}_m}}$ for $I_{i^{'}}\subseteq I_{i}$ suggests the following $J+1$ steps algorithm for $m=1,2$ to complete the optimal PMF vector $\mathbf{q}_[M]^{*}$:
\begin{itemize}
	\item Step $0$: For $\mu_{[MJ]}$ that does not satisfy $\mu_{m1}\geq\mu_{m2}\geq\cdots\geq\mu_{mJ_m}$, we can take a permutation $\Pi_m:\{1,\cdots,J_m\}\rightarrow\{1,\cdots,J_m\}$ such that $\mu_{m\Pi_m(i_m)}\geq\mu_{m\Pi_m(i_m+1)},i_m=1,\cdots,J_m-1$. Repeating Step $1$ to Step $J$, we can have similar result by substituting $\Pi_m(j)$ to $j$.
	\item Step 1: Assume $\mu_{m1}\geq\mu_{m2}\geq\cdots\geq\mu_{mJ_m}$. Since $\frac{\partial h(\mathbf{q})}{\partial q_{2^J-1}}\geq\frac{\partial h(\mathbf{q})}{\partial q_{i}}$ for all $i=0,1,\cdots,2^J-1$, $q^{*}_{m(2^J-1)}$ should be assigned the biggest allowable value. Note that $q_{m(2^J-1)}\leq\mu_{mj},j_m=1,\cdots,J_m$, we have
	\be 
	q^{*}_{m(2^J-1)}=\min\limits_{j_m=1,\cdots,J_m}\mu_{mj}=\mu_{mJ_m}.
	\ee 
	Due to constraints equation (\ref{eq.macqconstr}), we have $q^{*}_{mi_m}=0$ for $i_m$ with $b_{mJ_m}(i_m)=1$ and $i_m\neq2^{J_m}-1$.
	\item Step 2: For all $i_m\in\{0,1,\cdots,2^{J_m}-1\}$ and $b_{mJ_m}(i_m)=0$, we have $\frac{\partial h(\mathbf{q}_{[M]}}{\partial q_{i_m^{'}}}\geq\frac{\partial h(\mathbf{q}_{[M]})}{\partial q_{i_m}}$, where $i_m^{'}=\sum_{j_m=1}^{J_m-1}2^{j}_m-1$. Furthermore, for all $i_m$, $b_{mJ_m}(i_m)=0$, it follows that $q_{mi_m}\leq\mu_{mj}-\mu_{mJ_m}$ for $j_m=1,\cdots J_m-1$. Summarizing these facts, we have
	$q^{*}_{m(2^{J_m-1}-1)}=\mu_{J_m-1}-\mu_{J_m}$, and
	$q^{*}_{mi_m}=0$ for $i_m$ with $b_{mJ_m}(i_m)=0,b_{J_m-1}(i_m)=1$, and $i_m\neq\sum_{j_m=1}^{J_m-1}2^{j_m-1}$.
	\item Step $k_m$, $2<k_m<J_m$: Similar to Step $2$, we get $
	q^{*}_{m(2^{J_m-k_m+1}-1)}=\mu_{J_m-k_m+1}-\mu_{J_m-k_m+2}$ and 
	$q^{*}_{mi_m}=0$, for $i_m$ with $b_s(i_m)=0$,
	where $s=J_m-k_m+2,\cdots,J$, and $b_{m(J_m-k_m+1)}(i_m)=1,i_m\neq\sum_{j_m=1}^{J_m-k_m+1}2^{j_m-1}$.
	\item Step $J_m$: The only remaining PMF is $q_{m0}^{*}$ and $q_{m0}^{*}=1-\sum_{i_m=1}^{2^J_m-1}q_{mi_m}^{*}$.	
\end{itemize}
Thus, the right side of equation (\ref{eq.macjoint}) is maximized for $q_{mi}^{*}=\nu_{mi}$ if there exists $k_m\in\{0,1,\cdots,J\}$ such that $i_m=\sum_{j=1}^{k_m}2^{\Pi_m(j)-1}$; otherwise $q_i^{*}=0$, where
\be\label{eq.optnu2}
\nu_{mi}\dff\left\{\begin{array}{ll}
	1-\mu_{\Pi_m(1)},&i=0, \\
	\mu_{\Pi_m(i_m)}-\mu_{\Pi_m(i_m+1)},&i_m=1,\cdots,J_m-1, \\
	\mu_{\Pi_m(J_m)},&i=J_m;
\end{array}\right.
\ee 
and $\Pi_m:\{1,\cdots,J_m\}\rightarrow\{1,\cdots,J_m\}$ is a permutation of $\{1,\cdots,J_m\}$ such that $\mu_{\Pi_m(i_m)}\geq\mu_{\Pi_m(i_m+1)},i_m=1,\cdots,J_m-1$.
\subsection{Proof of Proposition \ref{prop.misomacwork2}}\label{appen.misomacwork2}
For simplicity, define $s_{mj_m}=p(\sum_{i_m=1}^{j_m}A_{mi_m}+\Lambda_0)$ for $m=1,2$, and $\hat{s}_{j_1j_2}=p(\sum_{m=1}^{2}\sum_{i_m=1}^{j_m}A_{mi_m}+\Lambda_0)$.
Based on symmetry, without loss of generality, assume $\mu_{m1}\geq\mu_{m2}\geq\cdots\geq\mu_{mJ_m}$. We know
\be
I_{MISO-MAC}(\mathbf{\mu}_{[MJ]})=h_b\Big(\sum_{j_1=0}^{J_1}\sum_{j_2=0}^{J_2}\nu_{1j_1}\nu_{2j_2}\hat{s}_{j_1j_2}\Big)-\sum_{j_1=0}^{J_1}\sum_{j_2=0}^{J_2}\nu_{1j}\nu_{2j}h_b\big(\hat{s}_{j_1j_2}\big),
\ee 
Note that $\sum_{j=0}^{J_m}\nu_{mj}=1$ for $m=1,2$, after rearrangement we get
\be 
&&I_{MISO-MAC}(\mathbf{\mu}_{[MJ]})\nonumber\\&=&h_b\Big(\sum_{j_2=0}^{J_2}\nu_{2j_2}\big[(1-\sum_{j_1=1}^{J_1}\nu_{1j_1})s_{2j_2}+\sum_{j_1=1}^{J_1}\nu_{1j_1}\hat{s}_{j_1j_2}\big]\Big)\nonumber\\&&-\Big(\sum_{j_2=0}^{J_2}\nu_{2j_2}\big[(1-\sum_{j_1=1}^{J_1}\nu_{1j_1})h_b(s_{2j_2})+\sum_{j_1=1}^{J_1}\nu_{1j_1}h_b\big(\hat{s}_{j_1j_2}\big)\big]\Big)\nonumber\\
&=&h_b\Big(\sum_{j_2=0}^{J_2}\nu_{2j_2}\big[s_{2j_2}+u_{2J_2}\big(\hat{s}_{j_1j_2}-s_{2j_2}\big)-\sum_{j_1=1}^{J_1-1}u_{1j_1}\big(\hat{s}_{j_1j_2}-s_{2j_2}\big)\big]\Big)-\sum_{j_2=0}^{J_2}\nu_{2j_2}\big[h_b(s_{2j_2})\nonumber\\&&+u_{2J_2}\big(h_b(\hat{s}_{j_1j_2})-h_b(s_{2j_2})\big)-\sum_{j_1=1}^{J_1-1}u_{1j_1}\big(h_b(\hat{s}_{j_1j_2})-h_b(s_{2j_2})\big)\big]
\ee 
where $u_{mj_m}\dff\nu_{mj_m}$ for $j_m=1,2,\cdots,J_m-1$, $m=1,2$ and $u_{J_m}\dff\sum_{j_m=1}^{J_m}\nu_{mj_m}$, then we have $0\leq u_{mj_m}\leq 1$ for $j_m=1,2,\cdots,J_m-1$ and $\max\{u_{mj_m},j_m=1,2,\cdots,J_m-1\}\leq u_{J_m}\leq 1$.
Note that $\hat{s}_{J_1J_2}=p(\sum_{m=1}^{2}\sum_{j_m=1}^{J_m}A_{mj}+\Lambda_0)\leq\frac{1}{2}$, then for $j_m=1,2,\cdots,J_m-1$,
\be 
\frac{\partial I_{MISO-MAC}}{\partial u_{2j_2}}&=&\underbrace{-h_b^{'}\Big(\sum_{j_2=0}^{J_2}\nu_{2j_2}\big[s_{2j_2}+u_{2J_2}\big(\hat{s}_{j_1j_2}-s_{2j_2}\big)-\sum_{j_1=1}^{J_1-1}u_{1j_1}\big(\hat{s}_{j_1j_2}-s_{2j_2}\big)\big]\Big)}_{\leq0}\nonumber\\&&\cdot\sum_{j_2}^{J_2}\nu_{2j_2}\big(\hat{s}_{j_1j_2}-s_{2j_2}\big)-\underbrace{\big[\big(h_b(\hat{s}_{j_1j_2})-h_b(s_{2j_2})\big)\big]}_{>0}<0
\ee 
and the max range of optimized $u_{J_m}$ corresponds to $u_{m1}=u_{m2}=\cdots=u_{J_m-1}$, we have the optimal $u_{[MJ]}$ satisfies $u_{m1}=u_{m2}=\cdots=u_{J_m-1}$, which implies $\mu_{m1}=\mu_{m2}=\cdots=\mu_{mJ}$ for $m=1,2$.

\renewcommand{\baselinestretch}{1.4}
\section{Auxilary Lemma}
\begin{lemma}\label{appenAu.convex}
	Assume function $f(x)$ is strictly convex and its first-order derivative exists. For $x>y$, then we have function $g(x,y)\dff\frac{f(x)-f(y)}{x-y}$ strictly monotonically increases with $x$, strictly monotonically decreases with $y$. To be specific, we have $f^{'}(y)<\frac{f(x)-f(y)}{x-y}<f^{'}(x)$ 
	\begin{proof}
		According to Lagrange mean value theorem, for $x>y$, we have $f(x)-f(y)=f^{'}(\xi)(x-y)<f^{'}(x)(x-y)$, where $y<\xi<x$. Since $g^{'}_x=\frac{f^{'}(x)(x-y)-[f(x)-f(y)]}{(x-y)^2}>0$, function $g(x,y)$ strictly monotonically increases with $x$. Similarly, we have function $g(x,y)$ strictly monotonically decreases with $y$.
		
		Note that function $g(x,y)$ strictly monotonically increases with $x$, we have $f^{'}(x)=\sup\limits_{y:x>y}\frac{f(x)-f(y)}{x-y}>\frac{f(x)-f(y)}{x-y}$ for any $y<x$. Similarly, we have $f^{'}(y)<\frac{f(x)-f(y)}{x-y}$.
	\end{proof}
\end{lemma}
\begin{lemma}\label{appenAu.hpconvex}
	Assume $\tau\leq\frac{\ln2}{b}$. $p(x)=1-e^{-x\tau}$, then we have function $h_b\big(p(x)\big)$, $x\in[0,b]$ is concave.
	\begin{proof}
		Note that $[h_b(p(x))]^{''}=h_b^{''}(p(x))[p^{'}(x)]^2+h_b^{'}(p(x))g^{''}(x)$, $h_b^{''}(x)<0$, $h_b(x)$ monotonically increase if $x\leq\frac{1}{2}$ and $p^{''}(x)<0$, we have $h_b^{''}\big(p(x)\big)<0$ when $\tau\leq\frac{\ln2}{b}$.
	\end{proof}
\end{lemma}
\begin{lemma}\label{appenAu.concavepro}
	Assume function $f(x)$ is concave. For $a<b<c<d$ and $a+d=b+c$, then we have $f(a)+f(d)<f(b)+f(c)$.
	\begin{proof}
		Note that $b-a=d-c$ and $f(x)$ is concave, then we have $f(b)-f(a)>f(d)-f(c)$.
	\end{proof}
\end{lemma}
\begin{lemma}\label{appenAu.convexyfrac}
	For $f(x)=\frac{d_2x+d_3}{d_1x+d_0}$, $x\in[0,1]$ and $\min\{d_0,d_1+d_0\}>0$. if $d_1<0$ and $d_2d_0-d_3d_1<(>)0$, then $f(x)$ is concave (convex) and monotonically decrease (increase).
	\begin{proof}
		 Taking the derivative of $f(x)$, we have $f^{'}(x)=\frac{d_2d_0-d_3d_1}{(d_1x+d_0)^2}$ and $f^{''}(x)=-2d_1\frac{d_2d_0-d_3d_1}{(d_1x+d_0)^3}$. Therefore, it is obvious to complete the proof.
	\end{proof}
\end{lemma}
\begin{lemma}\label{appenAu.last}
	For $a,b,c,d>0$, $a<c$, and $\frac{b}{a}<\frac{d}{c}$, then we have $\frac{b}{a}<\frac{d}{c}<\frac{d-b}{c-a}$ for any $\mu\in[0,1]$
	\begin{proof}
		It is easy to check
		\be 
		\frac{d}{c}<\frac{d-b}{c-a}\Leftrightarrow
		dc-da<dc-bc\Leftrightarrow bc<ad
		\ee
	\end{proof}
\end{lemma}
\begin{lemma}\label{appenAu.sisomacaux}
	For $a,b,c,d>0$, $a<c$, $\frac{b}{a}<\frac{d}{c}$ and $f(\mu)\dff \frac{\mu b+(1-\mu)d}{\mu a+(1-\mu)c}$, then we have $\mu\frac{b}{a}+(1-\mu)\frac{d}{c}<f(\mu)<\frac{d}{c}$ for any $\mu\in(0,1)$, $f^{'}(\mu)<0$ and $f^{''}(\mu)<0$
	\begin{proof}
    For the inequality on $f(\mu)$, we have
		\be 
		\frac{\mu b+(1-\mu)d}{\mu a+(1-\mu)c}>\mu\frac{b}{a}+(1-\mu)\frac{d}{c}\Leftrightarrow
		\frac{\mu(b-da/c)}{\mu a+(1-\mu)c}>\frac{\mu(b-da/c)}{a}\Leftrightarrow a<c, \nonumber
		\ee
	Based on Lemma \ref{appenAu.last}, we have
	\be 
	f^{'}(\mu)&=&c(\frac{d}{c}-\frac{d-b}{c-a})(c-a)[c-(c-a)\mu]^{-2}<0,\\
	f^{''}(\mu)&=&2c(\frac{d}{c}-\frac{d-b}{c-a})(c-a)^2[c-(c-a)\mu]^{-3}<0,
	\ee 
	\end{proof}
\end{lemma}

\renewcommand{\baselinestretch}{1.3}
\small{\baselineskip = 10pt
	\bibliographystyle{IEEEtran}
	\bibliography{maccap}
	
\end{document}